\documentclass[sigconf,screen]{acmart}
\usepackage{amsmath}

\usepackage{needspace}
\usepackage{algorithmic}
\usepackage{xcolor,listings}
\usepackage{graphicx}
\usepackage{url}
\usepackage{textcomp}
\usepackage{booktabs}
\usepackage{adjustbox}
\usepackage{longtable}

\usepackage[normalem]{ulem}

\usepackage{hyphenat}
\usepackage{cleveref}
\usepackage{multirow}
\usepackage{makecell}
\usepackage{enumitem}
\usepackage{colortbl}
\usepackage{tcolorbox}
\usepackage{supertabular}
\usepackage{hhline}
\usepackage{makecell}
\usepackage{caption}
\usepackage{subcaption}
\usepackage{tikz}
\usepackage[normalem]{ulem}
\usepackage[ruled,vlined]{algorithm2e}
\usepackage{pifont}
\usepackage[flushleft]{threeparttable}
\useunder{\uline}{\ul}{}
\setcopyright{none}
\renewcommand\footnotetextcopyrightpermission[1]{}

\newcommand\tool[1]{\emph{PyCPG}}
\newcommand\benchname[1]{\emph{\textit{PyVul}}}

\newcommand{\haowei}[1]{{\color{blue} #1}}

\definecolor{codegreen}{rgb}{0,0.6,0}
\definecolor{codegray}{rgb}{0.5,0.5,0.5}
\definecolor{codepurple}{rgb}{0.58,0,0.82}
\definecolor{backcolour}{rgb}{0.95,0.95,0.92}
\definecolor{nblue}{RGB}{165,11,210}

\lstdefinestyle{mystyle}{%
    keywordstyle=\color{nblue},%
    basicstyle=\ttfamily\footnotesize,%
    breakatwhitespace=false,%
    breaklines=true,%
    captionpos=b,%
    keepspaces=true,%
    numbers=left,%
    numbersep=5pt,%
    showspaces=false,%
    showstringspaces=false,%
    showtabs=false,%
    tabsize=2,%
}
\begin{document}

\title{An Empirical Study of Vulnerabilities in Python Packages and Their Detection}

\author{Haowei Quan}
\affiliation{%
  \institution{Monash University}
  \city{Melbourne}
  \country{Australia}}
\email{haowei.quan@monash.edu}
\author{Junjie Wang}
\affiliation{%
  \institution{College of Intelligence and Computing, Tianjin University}
  \city{Tianjin}
  \country{China}}
\email{junjie.wang@tju.edu.cn}
\author{Xinzhe Li}
\affiliation{%
  \institution{College of Intelligence and Computing, Tianjin University}
  \city{Tianjin}
  \country{China}}
\email{3022001754@tju.edu.cn}
\author{Terry Yue Zhuo}
\affiliation{%
  \institution{Monash University}
  \city{Melbourne}
  \country{Australia}}
\email{terry.zhuo@monash.edu}
\author{Xiao Chen}
\affiliation{%
  \institution{University of Newcastle}
  \city{Newcastle}
  \country{Australia}}
\email{xiao.chen@newcastle.edu.au}
\author{Xiaoning Du}
\affiliation{%
  \institution{Monash University}
  \city{Melbourne}
  \country{Australia}}
\email{Xiaoning.Du@monash.edu}

\begin{abstract}
In the rapidly evolving software development landscape, Python stands out for its simplicity, versatility, and extensive ecosystem. 
Python packages, as the unit for code organization, reusability, and distribution, have become an increasingly pressing concern, highlighted by the considerable number of vulnerability reports.
As a scripting language, Python often has to cooperate with other programming languages for usability, including better efficiency and interoperability with other libraries. 
This also adds complexity to the vulnerabilities inherent to Python packages, and the effectiveness of current vulnerability detection tools in spotting these vulnerabilities is underexplored within the research community.

This paper addresses these gaps by introducing \benchname{}, the first comprehensive benchmark suite of Python-package vulnerabilities.
\benchname{} includes 1,157 publicly reported, developer-verified vulnerabilities, each linked to its affected packages.
To accommodate diverse detection techniques, the benchmark provides annotations at both commit and function levels. 
An LLM-assisted data cleansing method is incorporated to improve label accuracy, achieving 100\% commit-level and 94\% function-level accuracy, establishing \benchname{} as the most precise large-scale Python vulnerability benchmark.
We further carry out a distribution analysis of \benchname{}, which demonstrates that vulnerabilities in Python packages involve multiple programming languages and exhibit a wide variety of types. 
Moreover, our analysis reveals that multi-lingual Python packages are potentially more susceptible to vulnerabilities.
Evaluation of state-of-the-art detectors using this benchmark reveals a significant discrepancy between the capabilities of existing tools and the demands of effectively identifying real-world security issues in Python packages.
Additionally, we conduct an empirical review of the top-ranked Common Weakness Enumerations (CWE) observed in Python packages, to diagnose the fine-grained limitations of current detection tools and highlight the necessity for future advancements in the field. 
\end{abstract}

\maketitle

\section{Introduction}

Over recent years, Python has become the leading programming language due to its user-friendly syntax, versatility, and rich ecosystem~\cite{ieee_spectrum}. With nearly 600,000 packages hosted on PyPI~\cite{pypi}, Python’s growing application across domains like web development and machine learning (ML) raises critical concerns about the security of its package ecosystem~\cite{ivaki2024taxonomy, tran2025detectvul}. For instance, web development often faces vulnerabilities such as cross-site request forgery (CSRF) and resource exhaustion, while ML packages are prone to issues like improper input validation. GitHub Advisory~\cite{githubadv} reported 507 vulnerabilities in Python packages in 2023, highlighting its growing security importance, comparable to npm (394) and Maven (937).



Despite this, no benchmark comprehensively captures real\hyp{}world Python package vulnerabilities with high accuracy. 
Real-world vulnerabilities in Python packages may involve other programming languages. Due to Python's nature as a scripting language, Python packages frequently embed C/C++ code for performance-critical tasks, such as the implementations of NumPy~\cite{NumPy} and PyTorch~\cite{PyTorch}.
Moreover, as a popular candidate for developing web applications, client-side code such as JavaScript and HTML can inevitably be involved.
This highlights the necessity of contextualizing vulnerabilities within Python packages rather than focusing solely on the Python code itself. Python packages provide a comprehensive view of execution paths that can reveal vulnerabilities and help reduce biases in vulnerability identification.

Current vulnerability benchmarks for Python, composed of vulnerabilities at either the commit or function level, either do not derive from or are difficult to associate with Python packages. For instance, CVEFixes~\cite{bhandari2021cvefixes} and CrossVul~\cite{nikitopoulos2021crossvul} are collected based on projects from security platforms such as National Vulnerability Database (NVD)~\cite{nvd} and do not effectively map to Python packages. In addition, datasets such as VUDENC~\cite{wartschinski2022vudenc} and SVEN~\cite{he2023large} focus on Python code changes, consequently overlooking cross-language vulnerabilities.
This motivates us to collect the first benchmark of real-world vulnerabilities in Python packages.

Concerns regarding the quality of existing benchmarks have been raised~\cite{croft2023data,chen2023diversevul}. In vulnerability-fixing commits, modified functions are often labeled as vulnerable, even when changes address non-vulnerability-related objectives like refactoring. This leads to inaccuracies in vulnerability assessments. 
Recent studies have attempted to address this challenge.
Wang et al.~\cite{wang2024reposvul} combined LLMs and static vulnerability detectors to determine vulnerable samples; however, 
as static vulnerability detectors are used for validating samples, the resulting dataset can no longer serve as a benchmark for them.
As revealed in our empirical assessment, current state-of-the-art static vulnerability detectors for Python suffer from low accuracy and excessive number of warnings, underscoring both their need for an effective benchmark and the inadequacy of relying on these detectors as validation mechanisms.
Ding et al.~\cite{ding2024vulnerability} proposed a heuristic-based labeling approach that significantly improves label accuracy but incurs substantial data loss and restricts the dataset to single-function vulnerabilities.
To overcome these issues, we make the first attempt to refine and cleanse the dataset using LLMs, with manual verification to ensure accuracy.

\setlength{\textfloatsep}{6pt} 
\begin{figure*}
    \centering
    \includegraphics[width=\textwidth]{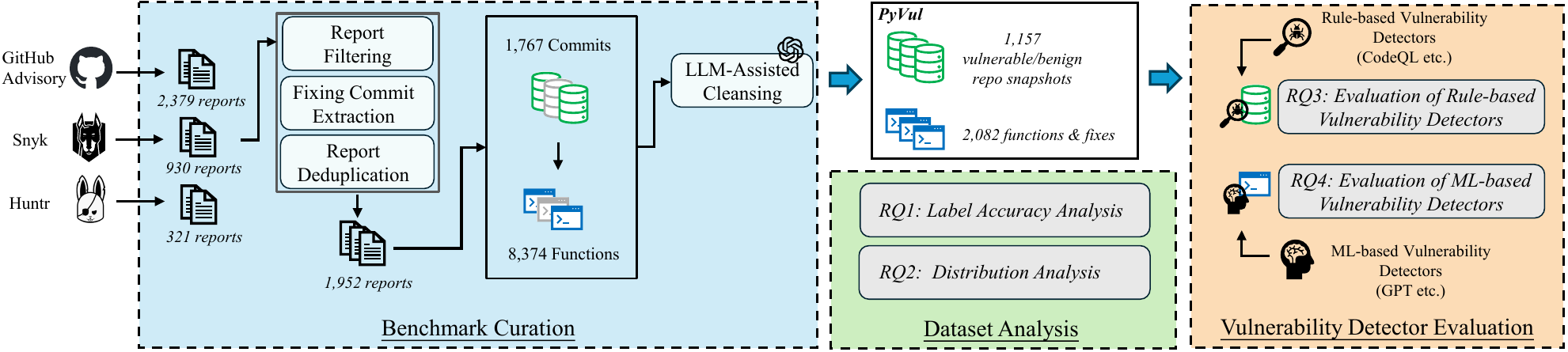}
    \vspace{-2em}
    \caption{The overview of our study}
    \label{fig:overview}
\end{figure*}

In this study, we present \benchname{}, \textbf{the first large-scale, high-quality vulnerability benchmark suite for Python packages.} 
We evaluate how effectively current vulnerability detectors can identify these vulnerabilities. 
\benchname{} consists of 1,157 verified vulnerabilities across 151 Common Weakness Enumeration (CWE) categories, identified in Python packages and refined through our LLM-based cleansing method, LLM-VDC.
To cater to the needs of vulnerability detectors operating at different granularities, 
we prepare our benchmark at both the commit level and the function level.
We initially collected 3,630 reports from three security advisories, GitHub Advisories~\cite{githubadv}, Snyk~\cite{snykio}, and Huntr~\cite{huntr}.
Subsequent filtering based on the existence of associated fixing commits and eliminating duplicates resulted in 1,767 reports. Following established methodologies from previous studies~\cite{bhandari2021cvefixes,nikitopoulos2021crossvul}, we constructed a commit-level benchmark consisting of 1,767 vulnerable repository snapshots, and a function-level benchmark comprising 8,374 vulnerable functions, each derived from the respective fixing commits.
To enhance the label accuracy of our benchmark, we developed and applied an LLM-assisted data cleansing method, \textit{LLM-VDC}. After cleansing, our benchmark, PyVul, achieves an accuracy rate of 100\% at the commit level with 1,157 repository snapshots, and an accuracy rate of 94.0\% at the function level with 2,082 vulnerable functions, as validated through random sampling.
This makes \textit{PyVul} 82.5\% to 92.8\% more accurate than previous automatically collected function-level datasets~\cite{bhandari2021cvefixes,nikitopoulos2021crossvul} and comparable to the human-annotated small dataset, SVEN~\cite{he2023large}, which contains only 380 vulnerable functions.
LLM-VDC demonstrates superior universality and outperforms the state-of-the-art labeling method introduced by Ding et al.~\cite{ding2024vulnerability} in 
a multi-lingual vulnerability dataset, with a 33.1\% greater improvement in function-level label accuracy.
After evaluating the effectiveness of our LLM-assisted data cleansing method and the quality of \textit{PyVul} (RQ1),
we analyze the distribution of real-world Python package vulnerabilities regarding programming languages, function compositions, and CWE categories (RQ2).

Furthermore, leveraging \textit{PyVul}, we 
assess state-of-the-art rule-based and ML-based static vulnerability detectors 
(RQ3 and RQ4). 
We leave the evaluation of dynamic vulnerability detectors to future work considering the common reproducibility issues in the open-source repositories~\cite{mukherjee2021fixing}.
The results of our evaluation show a significant gap in the ability of current Python static vulnerability detection tools to effectively report real-world security issues.
In addition to this assessment, we empirically review six most frequently reported CWEs in Python packages, aiming to provide insights into the limitations of current static tools and fuel future tools to detect zero-day vulnerabilities.
Our empirical study reveals significant discrepancies between the assumptions of current rule-based detectors and real-world security scenarios, compounded by a lack of support for 
most prevalent types of vulnerabilities such as those high-order vulnerabilities embedded in web applications, and a lack of support for Python's language features such as dynamic typing.
On the other hand, current ML-based detectors suffer from their unrealistic training data and function-level settings. Taking functions as input may result in models observing great variance in vulnerable samples or missing important context.

To summarize, our work makes the following contributions:
\begin{itemize}[leftmargin=*]
\item The first Python package vulnerability benchmark, \benchname{}, containing 1,157 commit\hyp{}level and 2,082 function-level vulnerabilities. 
It demonstrates an accuracy that is 82.5\% to 92.8\% higher than that of existing automatically collected function\hyp{}level vulnerability datasets, and it also excels in benchmark size.
\item An LLM-assisted approach, \textit{LLM-VDC}, for cleansing function-level vulnerability datasets, which demonstrates a 2.0 fold improvement in function-level label accuracy and enhances commit-level label accuracy to 100\%.
\item The first look into the distribution of Python package vulnerabilities regarding programming languages, function compositions, and CWEs.
\item A thorough evaluation of how well existing rule-based and ML-based detectors can identify vulnerabilities in \benchname{}, accompanied by an in-depth diagnosis of their major performance shortcomings for both approaches.
\item Our benchmark, code, and experimental scripts are made openly accessible at https://github.com/billquan/PyVul.

\end{itemize}

\section{Benchmark Construction}

In this section, we elaborate on the three steps used to establish a large and high-quality benchmark for Python package vulnerabilities, \benchname{}. These steps include data collection, benchmark curation, and data cleansing.

\subsection{Data Collection}
\begin{sloppypar}
We collect Python package vulnerabilities from three vulnerability reporting platforms that detail the ecosystems from which the vulnerabilities originate: GitHub Advisories~\cite{githubadv}, Snyk~\cite{snykio}, and Huntr~\cite{huntr}.
These platforms are widely used by developers and serve as data sources for other empirical research on vulnerabilities~\cite{zimmermann2019small,li2022mining,vasilakis2021preventing,bhuiyan2023secbench}.
In total, we gathered 2,379 vulnerability reports from GitHub Advisories, 930 from Snyk, and 321 from Huntr, totaling 3,630 reports. 
We conduct an initial screening, retaining those that meet the following criteria: 1) they include fix commits that address the corresponding vulnerabilities; 2) the repositories are accessible at the time of data collection, and the commits have not been rolled back or deleted; and 3) they are not duplicates of any other reports. 
Ultimately, our initial collection resulted in 1,767 unique vulnerability reports.
\end{sloppypar}

\subsection{Benchmark Curation}\label{subsec:curation}

A comprehensive and accurate benchmark for vulnerabilities is essential for evaluating vulnerability detectors. Currently, there are two main categories of static vulnerability detection methods applicable to Python packages: rule-based and ML-based methods.
These methods differ significantly in terms of the context granularity they use for identification. Rule-based static analysis methods, such as CodeQL~\cite{CodeQL}, PySA~\cite{PySA}, and Bandit~\cite{Bandit}, typically operate at the project level. In contrast, ML-based static analysis methods~\cite{chakraborty2021deep, chen2023diversevul, ding2024vulnerability} generally work at the function level.
To cater the needs of both types of analysis, we have constructed a benchmark that accommodates vulnerabilities at both the commit  and function levels.


\textbf{Commit-level Benchmark.}
To construct the commit-level benchmark, we checkout the 1,767 collected commits as patched, non-vulnerability samples and their direct parent version on the main branch as vulnerability samples. 


\textbf{Function-level Benchmark.}
To construct the function-level vulnerability dataset, we employed a common methodology utilized in previous studies ~\cite{chakraborty2021deep, chen2023diversevul, ding2024vulnerability}. For each commit, we consider the functions involved as vulnerability samples in their pre-fix version and as non-vulnerability samples in their post-fix version. From a total of 1,767 commits, we collected 8,374 vulnerability samples and 8,374 non-vulnerability samples, resulting in a comprehensive dataset of 16,748 samples.



\subsection{Data Cleansing}
To evaluate the quality of the curated benchmark and compare it with baseline benchmarks, we manually validated a statistically significant number of randomly sampled vulnerable commits and vulnerable functions from \benchname{}, \textit{CVEFixes}~\cite{bhandari2021cvefixes}, and \textit{CrossVul}~\cite{nikitopoulos2021crossvul} in Section~\ref{subsec:label_acc}. The sample sizes were determined following~\cite{william1977sample}, and 
the results are summarized in Table~\ref{tab:label_acc}.
The accuracy of commit-level labels significantly surpasses that of function-level labels \linebreak across all three benchmarks, achieving rates of 99.7\%, 99.5\%, and 99.4\%, respectively. 
On the other hand, the function-level label accuracy stands at only 40.4\%, 48.3\%, and 51.0\% for the three benchmarks. The low quality arises mainly because numerous changes in these commits do not pertain directly to vulnerabilities; instead, they involve code refactoring, consistent code style maintenance, or improvements in code readability, which are also highlighted by other researchers~\cite{ding2024vulnerability}. 
For developers with security background, pinpointing the actual vulnerability-fixing changes within a commit is not notably difficult. However, manually annotating all samples within the benchmark is labor-intensive and not scalable. As an alternative, we propose an approach, \textit{LLM-VDC}, that leverages the code semantic understanding capabilities of LLMs to help filter out function-level changes that are unrelated to vulnerability fixes. 
This approach additionally improves the commit-level label accuracy, as we only retain the commits with at least one relevant function change.

We utilize the in-context learning capabilities of LLMs due to the insufficient fine-tuning data available for our annotation task. We implement established practices of prompt engineering when presenting the task to LLMs. These practices include system role definition~\cite{zheng2023helpful}, few-shot learning~\cite{brown2020gpt}, and chain-of-thought (CoT) prompting~\cite{wei2022chain}, all of which have proven effective in recent studies.
The system role defines how LLMs should function during interactions, influencing the tone, focus, and limitations of their responses. In this case, we designate the system role as a security expert.
Few-shot learning enables LLMs to grasp and perform specific tasks through illustrative examples, while the CoT prompting technique enhances these examples by outlining the reasoning process behind the answers.
In our design, we specifically ask LLMs to state the reasons before yielding the final answer.

\setlength{\textfloatsep}{10pt}
\begin{figure}[tbp]
    \centering
    \includegraphics[width=\linewidth]{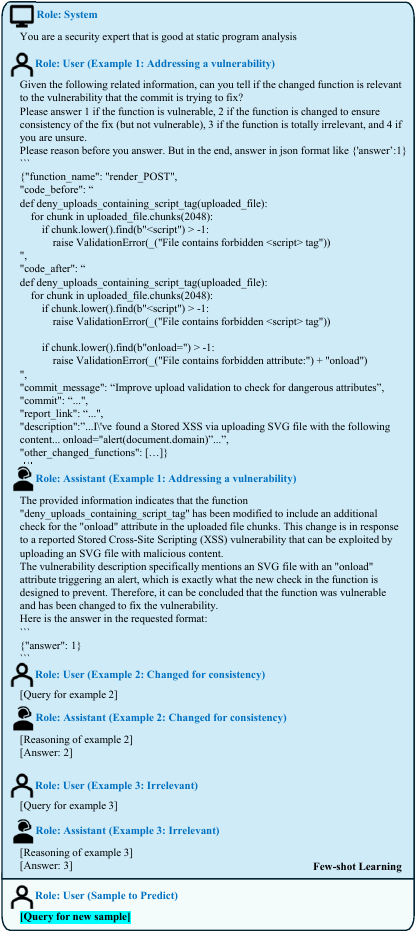}
    \caption{The prompt used for annotating the relevance of function-level changes to vulnerability fixes. 
    }
    \label{fig:llm_filtering_workflow}
\end{figure}

We now elaborate on the task formulation, emphasizing the clarity and inclusiveness of classification categories, the adequacy of context information, and the adaptive truncation of context to accommodate the token limitations of LLMs. 
\begin{itemize}[leftmargin=*]
\item \textbf{Adequacy of Context.} 
The task involves determining whether the changes made to a function are intended to address security issues identified in the associated commit. To assist in this assessment, we provide two main pieces of information: 1) Details about the focal function, including its name and complete function definition both before and after the changes, and 2) Contextual information regarding the vulnerability being addressed, which includes the commit message, a link to the advisory report, a description of the vulnerability from the report, and a list of all other functions that were modified in this commit.

\item \textbf{Classification Categories.} 
To help LLMs better understand our task, we explicitly ask the LLMs to classify each change into one of four categories: 1) the function is patched against a vulnerability; 2) the function is not vulnerable but has been changed for consistency; 3) the function is irrelevant to the vulnerability; or 4) no decision can be made.
\textbf{We provide LLMs with one example for each of the first three categories through few-shot learning, accompanied by detailed reasoning steps} elucidating why the example aligns with the respective category.
This method not only aids LLMs in accurately understanding the category definitions but also in adopting the intended reasoning processes.
It is important to note that LLMs have the option to indicate when they cannot reach a clear conclusion. This feature helps prevent the model from generating hallucinations by avoiding incorrect answers when it is uncertain.
\haowei{
}

\item \textbf{Adapt to Token Limitations.}
Due to the limited context length of LLMs, we implement strategies to sacrifice part of the context information when the limit is exceeded. Specifically, we employ the following measures:
1) Commit messages or descriptions from advisory reports will be truncated with a note stating ``collapsed due to token limitation'' if they exceed a certain threshold (e.g., 2,000 characters, as used in our experiment), and 2) We adopt a stepwise reduction method to supply information about other changed functions in the commit.
Depending on whether the context limitation of the prompt has been exceeded, we will attempt the following methods from \ding{182} to \ding{184} and only choose the latter option if all former options are infeasible:
\ding{182} We supply all other changed function information in the commit;  
\ding{183} We supply all other changed functions in the same file;  
\ding{184} We do not supply other changed functions.  
\end{itemize}

The prompt used to annotate the relevance of each modified function with the vulnerability-fixing commit is shown in Figure~\ref{fig:llm_filtering_workflow}.
In this study, we adopt GPT-4~\cite{GPT}, one of the top-performing LLMs available at the time of writing.
Our cleansing method results in a collection of 1,157 commits and 2,082 function pairs.


\section{Study Design}
In this study, we aim to investigate the characteristics of vulnerabilities in Python packages and assess the performance of current static vulnerability detection methods on these vulnerabilities. We begin by evaluating the label accuracy of \benchname{} and use \benchname{} as a foundation for our analysis in the following sections. The evaluation focuses on addressing the following research questions (RQs):
 \begin{itemize}[leftmargin=*]
 \item RQ1: How accurate are the vulnerability labels in \benchname{}?
 \item RQ2: What is the vulnerabilities distribution in Python packages?
 \item RQ3: How effective are current rule-based approaches for detecting vulnerabilities in \benchname{}?
 \item RQ4: How effective are current ML-based approaches for detecting vulnerabilities in \benchname{}?
 \end{itemize}


Next, we will introduce the subjects used for comparison and evaluation when addressing these RQs. This includes existing benchmarks for Python vulnerabilities, automated filtering methods for vulnerability datasets, rule-based and ML-based vulnerability detectors that are relevant to the Python community.

\subsection{Existing Python Vulnerability Benchmarks}
To the best of our knowledge, there is currently no dataset specifically focused on vulnerabilities in Python packages. However, several datasets have been curated that concentrate on vulnerabilities within Python programs. 
In this study, we utilize the label accuracy of these existing datasets to benchmark the label accuracy of \benchname{}. We select the subject datasets based on two criteria:
1) the dataset must contain vulnerable Python code; and
2) the vulnerabilities in the dataset should either be manually verified or linked to corresponding advisory reports or Common Vulnerabilities and Exposures (CVE) entries, ensuring the dataset's high quality.


\begin{itemize}[leftmargin=*]
\item \textit{SVEN}~\cite{he2023large}. 
SVEN is a manually annotated vulnerability dataset that contains 808 pairs of vulnerable and non-vulnerable functions across various programming languages. Among these, 380 pairs refer specifically to Python functions. SVEN was con-\linebreak structed by carefully examining existing data sources, including \linebreak VUDENC~\cite{wartschinski2022vudenc} and BigVul~\cite{fan2020ac}, to ensure high quality. However, the intensive manual effort involved in creating this dataset limits its scalability.
It is important to note that the original SVEN dataset does not include commit-level vulnerabilities, which are essential for rule-based vulnerability detectors that operate on repository snapshots. To address this gap, we enhanced the SVEN dataset by incorporating the relevant fix commits for the identified vulnerable and non-vulnerable function pairs. This supplementation resulted in a total of 143 commit-level vulnerability and non-vulnerability samples.

\item \textit{CVEFixes~\cite{bhandari2021cvefixes} and CrossVul~\cite{nikitopoulos2021crossvul}.}
Both CVEFixes and CrossVul are multi-language datasets derived from the CVE database, which include the fixing commits for various vulnerabilities. 
CVEFixes is annotated at the function level, whereas CrossVul provides annotations at the file level.
Previous research~\cite{chen2023diversevul} has successfully extracted vulnerable and benign functions from the CrossVul dataset. We employ a similar approach to obtain function-level data for CrossVul. 
As a result, we have 1,360 pairs of Python functions related to 508 commits in CVEFixes, and 777 pairs of Python functions related to 319 commits in CrossVul.
\end{itemize}


\subsection{Vulnerability Dataset Cleansing Methods}

The development of effective cleaning approaches and highly accurate techniques for annotating vulnerable functions has been limited. To our knowledge, the most advanced automated labeling method is presented by Ding et al.~\cite{ding2024vulnerability}. 
They proposed a set of heuristic rules to retain functions only if the likelihood of them being the source of a vulnerability is high.
These heuristic rules are heavily focused on the precision of identifying vulnerable functions, rather than on recall. Specifically, a prior version of a function is labeled as vulnerable when it meets all three of the following criteria:
1) the function is the only one modified in a vulnerability-fixing commit;
2) the function's name is mentioned in the linked CVE report's vulnerability description; and
3) the function's file name is noted in the associated CVE report's vulnerability description, and it is the only function modified in that file.
Using this approach, PrimeVul achieves a high accuracy rate of around 90\%. However, its strict rules may lead to a significant loss of vulnerable samples, as many vulnerabilities are not limited to a single function. Additionally, CVE descriptions can be incomplete or may not always specify particular functions or file names. 
In contrast, our data cleaning method, \textit{LLM-VDC}, is more adaptable and can accurately identify genuine vulnerable functions across a wide range of commits with the help of LLMs. 
In the following, we will use the name \textit{PrimeVul} to also refer to this rule-based method for cleansing vulnerability datasets.

\subsection{Static Vulnerability Detectors}

Below, we introduce the three rule-based approaches and three machine learning-based approaches used to analyze their effectiveness in detecting vulnerabilities in \benchname{}. 

\subsubsection{Rule-based Approaches.} 

Our selection criteria are as follows: the methods must support vulnerability detection for Python, be executable, and be widely recognized in the field of vulnerability detection, reflecting the highest standard in this task.

\begin{itemize}[leftmargin=*
]
\item \textit{CodeQL}~\cite{CodeQL}.
CodeQL is a comprehensive static analysis engine developed by GitHub that uses queries to identify vulnerable patterns in code. It converts the source code of a program into a queryable database that maintains the program's semantics, such as data and control flows. Additionally, CodeQL comes with a suite of query-based rules designed to detect various types of vulnerabilities.
CodeQL supports multiple programming languages, including Python. 
For Python, there are 101 built-in \linebreak queries~\cite{codeqlrules}, including an extended set focused on security. Each query is annotated with the CWEs that it aims to detect, and in total, 123 CWEs are covered by these queries. 

\item \textit{PySA}~\cite{PySA}. 
PySA is a static code analysis tool developed as part of Facebook's Pyre-check project. It is specifically designed to address a variety of taint-style vulnerabilities in Python applications, such as SQL injections. 
PySA identifies and flags these vulnerabilities by analyzing the code to trace data flows from untrusted input sources to potentially vulnerable sinks. The tool features 38 clearly defined taint analysis rules that describe the characteristics of the sources and sinks
, which help in identifying covered CWEs. 
In total, PySA covers 67 different CWEs.

\item \textit{Bandit}~\cite{Bandit}. 
Bandit is a popular open-source static analysis tool (with 6k stars on GitHub) designed specifically to identify security issues in Python code. 
It scans Python programs to uncover common vulnerabilities, such as the use of potentially dangerous APIs and hard-coded credentials. 
Bandit focuses on detecting vulnerabilities primarily through pattern matching against Abstract Syntax Trees (ASTs) and does not take into account control flows or data flows within the code.
It encompasses 39 rules, each annotated with the CWE related to the target vulnerability. In total, these rules cover 17 distinct CWEs.
\end{itemize}
\vspace{-1em}
\subsubsection{ML-based Approaches.} 
\label{sec:ml-approaches}
The state-of-the-art ML-based ap- \linebreak proaches for Python vulnerability detection primarily include GNN-based and LLM-based methods.
GNN is a widely used model architecture for vulnerability detection. To the best of our knowledge, the state-of-the-art GNN method trained to detect Python vulnerabilities is VUDENC~\cite{wartschinski2022vudenc}. However, due to unresolved bugs in its published implementation~\cite{vudencrepo} and our unsuccessful attempts to contact the authors, we were unable to replicate VUDENC and therefore had to exclude it from our study. 



LLMs pretrained on code have significant potential for vulnerability detection, either through direct prompting~\cite{ding2024vulnerability} or fine- tuning~\cite{chen2023diversevul, ding2024vulnerability}. Research indicates that models from the GPT-2 family can achieve performance comparable to GNN-based methods when fine-tuned on small datasets, such as CVEFixes~\cite{bhandari2021cvefixes}, and demonstrate superior performance on larger datasets~\cite{chen2023diversevul}.
However, due to a lack of specifically curated Python datasets, most existing studies have focused on other programming languages, particularly C/C++. 
In this paper, using \benchname{}, we conduct the first study of LLM-based vulnerability detection in Python packages.
In RQ4, we will evaluate the performance of LLMs employing two distinct methods: 1) direct prompting, which assesses the LLM’s inherent knowledge of vulnerabilities and its ability to identify vulnerable code patterns, and 2) fine-tuning, which examines the models' capacity to learn from vulnerability samples and adapt to the task of vulnerability detection. 
For our experiments, we selected three LLMs, including one open-source model and two from OpenAI.

\begin{itemize}[leftmargin=*
]
\item \textit{CodeQwen1.5-Chat}~\cite{CodeQwen1.5}. 
CodeQwen1.5-Chat is one of the latest code LLMs from the open-source community, featuring 7B parameters. This model has been pretrained on approximately 3 trillion tokens of code-related data. We have selected CodeQwen1.5-Chat, the instruction-tuned version of CodeQwen1.5, to evaluate its performance in both direct prompting and fine-tuning scenarios. Despite having only 7B parameters, CodeQwen1.5-Chat has demonstrated state-of-the-art performance on the HumanEval benchmark~\cite{humaneval}, outperforming GPT-3.5 Turbo and showing performance comparable to GPT-4.
\item \textit{GPT-3.5 Turbo} and \textit{GPT-4}~\cite{GPT}. 
The GPT family of models demonstrates exceptional performance among LLMs and shows superior capabilities compared to open-source models in vulnerability detection~\cite{ding2024vulnerability}. As of this study, GPT-3.5 Turbo is the highest-performing proprietary model available for fine-tuning, while GPT-4 ranks as the top-performing proprietary model on the HumanEval leaderboard.

\end{itemize}

\subsubsection{Experimental Setup.}
\label{sec:exp-setup}
When evaluating the rule-based ap- \linebreak proaches, we used their default settings. Due to time constraints, we set a timeout of 60 minutes for each run of the detectors. 
All evaluations of the rule-based approaches were conducted on a computer equipped with a 14-core Intel Xeon W-2175 CPU and 32 GB of RAM, running Ubuntu 20.04.6.


We follow existing research~\cite{ding2024vulnerability} to setup the fine-tuning framework for CodeQwen1.5-Chat, with Axolotl~\cite{axolotl}.
We load the model's weights from Hugging Face Models~\cite{huggingface} and fine-tune it with a learning rate of $2 \times 10^{-5}$ for four epochs using LoRA~\cite{hu2021lora}, which balances between the training efficiency and task performance~\cite{zhuo2024astraios,weyssow2023exploring}. 
To directly prompt CodeQwen1.5-Chat, we utilize the default parameters, and all experiments are conducted on an NVIDIA A100 GPU with 80 GB of memory. 
For the proprietary models, we interact with them through the OpenAI APIs. 
We fine-tune GPT-3.5 Turbo for just one epoch following~\cite{ding2024vulnerability}. Since fine-tuning for GPT-4 was not available during this study, our evaluation does not include this model. 
During the model inference, we use the models' default parameters while setting the temperature parameter to 0 for deterministic results.

\section{Experiment Results}
\subsection{RQ1: Benchmark Quality}\label{subsec:label_acc}


\begin{table*}[tbp]
\centering
\footnotesize
\caption{Label accuracy of existing benchmarks and our newly curated benchmark, and the impact of data cleansing method on label accuracy. 
}
\setlength{\tabcolsep}{4pt}
\vspace{-0.5em}
\label{tab:label_acc}
\centering
\begin{tabular}{lccccccccc}
\toprule
Benchmark &\makecell{Data Cleansing}&  \#Commits  & \makecell{\#Sampled\\ Commits}&\makecell{Commit-lvl \\Acc(\%)}&\makecell{\#Vuln Funcs}  & \makecell{\#Sampled \\Funcs}& \makecell{Function-lvl\\ Acc(\%) }\\
\midrule
SVEN\cite{he2023large} &N/A&  143 &105&100&380 &192& 96.3 \\
CVEFixes\cite{bhandari2021cvefixes} &N/A& 508 &219&99.5 & 1,360 &300& 48.3 \\
CrossVul\cite{nikitopoulos2021crossvul} &N/A& 319 &175&99.4& 777 &258& 51.0 \\
\textbf{PyVul (Python)} &\textbf{LLM-VDC}& \textbf{788} &259&\textbf{100.0}& \textbf{1,480} &306& \textbf{93.1} \\
\hline
     &N/A&1,767&316&99.7& 8,374 &368& 40.4 \\
\textbf{PyVul} &PrimeVul&745&254&100.0& 1,012 &279& 70.8 \\
 &\textbf{LLM-VDC}&\textbf{1,157}&289&\textbf{100.0}&   \textbf{2,082} &325& \textbf{94.2} \\ 
\bottomrule
\end{tabular}
\end{table*}

\subsubsection{Effectiveness of LLM-VDC}
\label{sec:llm-vdc-eval}
To assess the effectiveness of our LLM-assisted method for cleansing vulnerability datasets, we evaluate the label accuracies of \benchname{} before any data cleansing, after applying \textit{PrimeVul}, and after applying \textit{LLM-VDC}, respectively.

Validating the raw data, which includes 1,767 commits and 8,374 functions, is not feasible to do manually. 
Therefore, we evaluate the data by randomly sampling commits and vulnerable functions from each version of the dataset and manually checking the accuracy of their labels. The sample sizes $n_{sample}$ are computed by the formula~\cite{william1977sample} for statistical significance, $n_{sample} =  \frac{z^2\times p(1-p)/e^2}{1+z^2\times p(1-p)/e^2N}$, where $N$ is the population size, $z$ is the z-score, $p$ is the standard deviation, and $e$ is the margin of error. We adopt a z-score of 1.96 corresponding to 95\% confidence level, a standard deviation of 0.5, which is the maximum standard deviation as the exact distribution unknown, and a margin of error of 5\%.
Columns four and seven in Table~\ref{tab:label_acc} lists the sample sizes.
The first and the third authors, both with expertise in general software vulnerabilities and over four years of experience in Python development, independently assess whether each commit or function is genuinely related to the reported vulnerability. Their evaluation is based on a thorough review of the fix commit, the code before and after the commit, the related vulnerability report, the CVE descriptions, and any available discussions among developers.
In cases where the authors disagree, they collaboratively review the sample and discuss it until they reach a consensus.
For our assessment criteria, a commit is considered erroneously labeled if: 1) it is irrelevant to the associated vulnerability, or 2) it does not fully resolve the vulnerability. A function is deemed mistakenly labeled as vulnerable if: 1) it has been modified for consistency rather than directly addressing the vulnerability, or 2) it is entirely irrelevant to the vulnerability. 
We calculate label accuracy as the percentage of correctly labeled samples out of all evaluated samples. The results of this analysis are presented in Table~\ref{tab:label_acc}.


When no data cleansing method is applied, the collected dataset shows a label accuracy of 99.7\% at the commit level. Both \textit{PrimeVul} and \textit{LLM-VDC} successfully increase the commit-level label accuracy to 100.0\%.
At the function level, when no data cleansing is applied, the label accuracy of the collected raw data is 40.4\%. After applying the \textit{PrimeVul} method, the label accuracy increases to 70.8\%. However, this is significantly lower than the accuracy of around 90\% reported in their own datasets~\cite{ding2024vulnerability}. This indicates that \textit{PrimeVul} shows limited effectiveness in our \benchname{}.
The primary reason for its underperformance is presumably the inclusion of diverse programming languages, particularly the obfuscated JavaScript code. 
\textit{PrimeVul} relies on searching for function names in the CVE descriptions, but changes in obfuscated JavaScript code often involve short, non-descriptive function names, such as a single letter ``p''. 
These functions are likely to be inaccurately associated with vulnerability fixes.
This highlights \textit{PrimeVul}'s limitations in generalization ability.
In contrast, the \textit{LLM-VDC} method significantly enhances the label accuracy of \benchname{} to 94.2\%.

Additionally, \textit{LLM-VDC} retains a significantly larger number of samples compared to \textit{PrimeVul}. Initially, the benchmark shows 1,767 commits and 8,374 vulnerable functions before any data cleansing. After applying \textit{PrimeVul}, these numbers decrease to 745 commits and 1,012 functions. In contrast, the \textit{LLM-VDC} method retains 1,157 commits and 2,082 functions. Remarkably, at the function level, \textit{LLM-VDC} retains twice as many samples as \textit{PrimeVul}, highlighting its superior effectiveness in data preservation.


\begin{tcolorbox}[left=6pt,right=6pt,boxsep=-1mm]
\textit{LLM-VDC} significantly enhances the function-level label accuracy of \benchname{} to 94.2\%. Compared to the baseline cleansing method, \textit{LLM-VDC} not only achieves a greater improvement in label accuracy but also retains twice as many samples.
\end{tcolorbox}

\subsubsection{Label Accuracy of \benchname{}}
Since both \textit{PyVul} and the baseline benchmarks include code written in Python and other programming languages, we extracted and evaluated only the segments pertinent to Python code to ensure a fair comparison. We refer to this subset of our \textit{PyVul} benchmark as \textit{PyVul (Python)}. 
In this section, we randomly sample commits and vulnerable functions from each benchmark and follow the same procedure for manually validating their label accuracy as described in Section~\ref{sec:llm-vdc-eval}.

We present the results in Table~\ref{tab:label_acc}. As indicated, all four benchmarks demonstrate strong performance in commit-level labeling accuracy, achieving 99.4\% or higher.
At the function level, we observe that the label accuracy of the automatically collected vulnerability datasets, CVEFixes and CrossVul, is 48.3\% and 51.0\%, respectively. In contrast, the manually curated vulnerability function dataset, SVEN, boasts a label accuracy of as high as 96.3\%. However, this dataset is limited in size due to the high cost of manual annotation, containing only 380 vulnerable functions.
On the other hand, the vulnerability function dataset we collected automatically has a label accuracy of 93.1\%. 
This accuracy is between 82.5\% and 92.8\% higher than that of the other two automated baseline benchmarks and is comparable to the manually curated dataset SVEN.
With the advanced data cleansing method, opportunities arise to establish a high-quality vulnerability benchmark on a large scale.

\begin{tcolorbox}[left=6pt,right=6pt,boxsep=-1mm]
All the Python vulnerability benchmarks examined demonstrate high accuracy at the commit level. However, the accuracy at the function level varies significantly. 
\textit{PyVul (Python)} achieves a label accuracy of 93.1\%, which is between 82.5\% and 92.8\% higher than the accuracies of the automatically collected baseline datasets. Additionally, this accuracy is comparable to that of the manually curated dataset SVEN.

\end{tcolorbox}

\setlength{\floatsep}{8pt}
\begin{table}[tbp]
\centering
\footnotesize
\caption{The language composition of \benchname{}.}
\vspace{-0.5em}
\label{tab:pyvul_language}
\begin{tabular}{crr}
\toprule
\textbf{Language Composition} &  \textbf{\#Commits} & \textbf{\#Functions} \\
\midrule
Python & 775 (67.0\%) & 1,480 (71.1\%) \\
C/C++ & 335 (29.0\%) & 463 (22.2\%)\\
JavaScript/TypeScript & 23 (2.0\%) & 115 (5.5\%)\\
Java & 4 (0.3\%) & 12 (0.6\%) \\
Other Language & 2 (0.3\%) & 12 (0.6\%) \\ 
Multiple Languages &  18 (1.6\%) & -\\
\midrule
\textbf{Total}  & 1,157 & 2,082 \\
\bottomrule
\end{tabular}
\end{table}

\subsection{RQ2: Characteristic Analysis of Python Package Vulnerabilities}\label{subsec:vuln_distribution}


Currently, there is no systematic analysis of the characteristics of Python package vulnerabilities due to the absence of a vulnerability benchmark for Python packages. 
The curated \textit{PyVul} benchmark allows for a comprehensive analysis of various aspects of Python package vulnerabilities. This includes qualitative and quantitative analysis of their language composition, the number of functions involved, and the types of vulnerabilities present, providing insights necessary for understanding Python package vulnerabilities and guiding the development of corresponding detection tools.



\subsubsection{Language Composition Analysis}


The analysis of language composition in the benchmark offers valuable insights for developing effective detection tools. 
Since packages from which the vulnerabilities originate provide a more comprehensive context for understanding them, while the fixing commits directly indicate their causes and fixes, our analysis is conducted against both.

\begin{figure}[tbp]
    \centering
    \includegraphics[width=\linewidth]{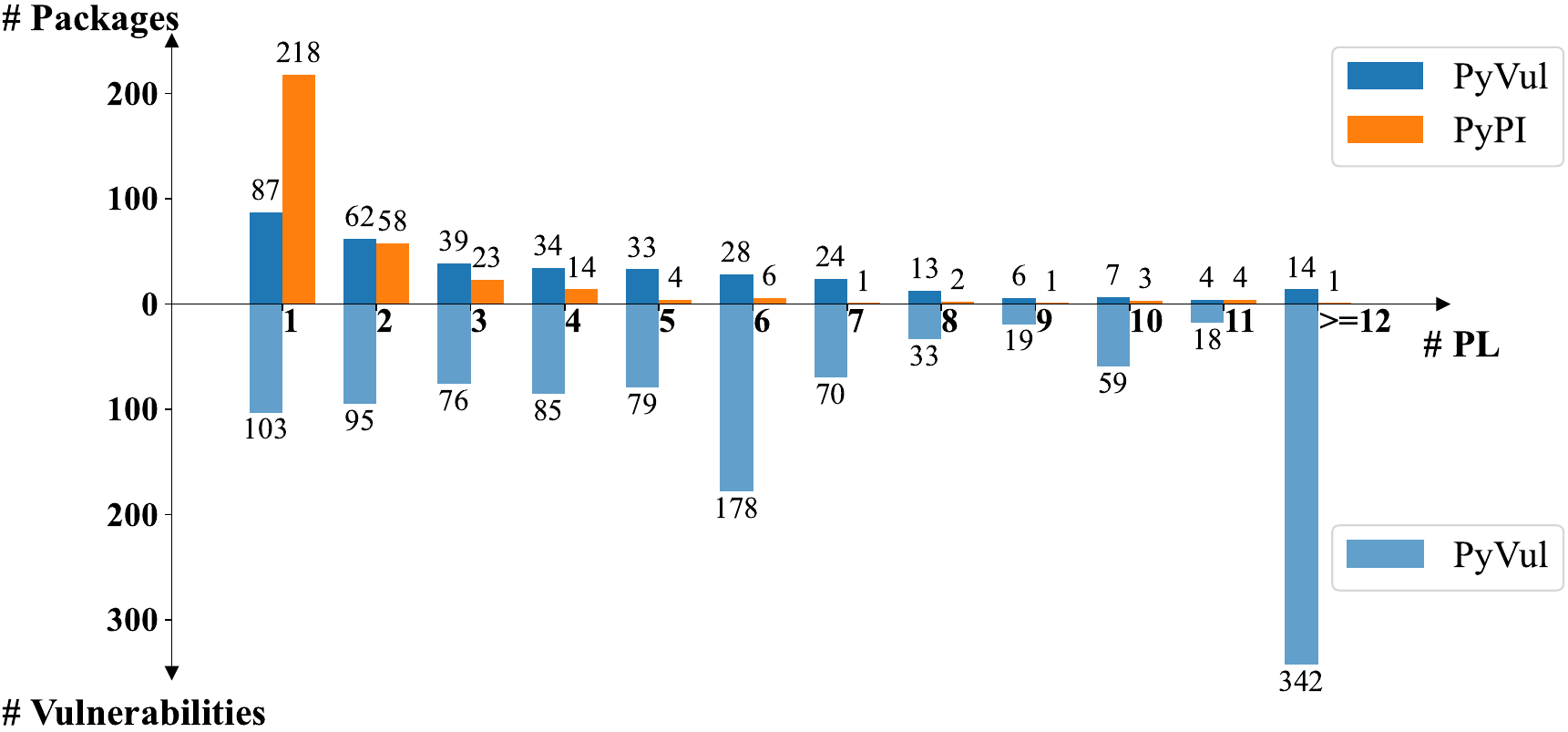}
    \caption{Programming language (PL) distribution in Python packages.}
    \label{fig:pkg_composition_code_change}
\end{figure}

We conducted an analysis of programming languages used in all 349 Python packages associated with \benchname{} by querying the language statistics of their repositories via the GitHub API.
The results are presented in Figure~\ref{fig:pkg_composition_code_change}. 
As shown, these Python packages predominantly involve multiple programming languages. Approximately 75\% (262/349) of the packages used at least two programming languages, while around 36\% (127/349) utilized at least five different languages. 
We additionally counted the total number of vulnerabilities at the commit level encompassed by these packages as light blue bars in Figure~\ref{fig:pkg_composition_code_change}.
The data reveals that over 90\% of the vulnerabilities are found in packages that use multiple languages. 
On average, a Python-only package is associated with 1.18 vulnerabilities, while a multi-lingual Python package is linked to 4.02 vulnerabilities.
Notably, 14 packages with more than 12 languages contribute 342 vulnerabilities. The main reason is that two of these packages, TensorFlow and PyTorch, account for 311 vulnerabilities.
We further employ the interquartile range method~\cite{dekking2006modern} to remove the impact of outliers. After adjustment, a Python-only package is associated with an average of 1.18 vulnerabilities, whereas a multi-lingual Python package averages 1.29 vulnerabilities.
This suggests an increased risk of vulnerabilities in multilingual packages.

To better understand the relationship between vulnerabilities and the multi-language characteristics of Python packages, we compare the language composition distribution of the packages in \benchname{} with that of general PyPI packages. 
The packages in \benchname{} are quite popular, averaging 13,358.7 stars on GitHub.
To effectively control the effect of popularity, we randomly select the same number of packages from the top 8,000 most popular PyPI packages~\cite{pypipop} for comparison. 
As illustrated in Figure~\ref{fig:pkg_composition_code_change}, the packages in \benchname{} show a clear tendency towards the usage of multiple programming languages. 
This echoes the observation that multi-lingual Python packages can be more susceptible to vulnerabilities, which is also consistent with the observation in previous work~\cite{yang2024multi}.

We further analyze the language composition of vulnerabilities at the commit level. 
We use Guesslang~\cite{guesslang} to identify each vulnerable function’s programming language, and aggregate them to derive the language composition at the commit level.
We group the commits and vulnerable functions, respectively, according to their programming languages and present the statistics in Table~\ref{tab:pyvul_language}. 
To our surprise, only 1.6\% of the vulnerabilities involve more than one programming language. 
Among the vulnerabilities, 67.0\% are exclusively related to Python, while 31.4\% are associated with other programming languages, with C/C++ being the most prevalent non-Python language.
Two important observations can be drawn: 1) non-Python vulnerabilities are common in Python packages, and 2) most vulnerabilities and their fixes are associated with a single programming language. 
It is essential to note that this does not imply that they can be effectively detected by tools designed for that specific language. 
The broader context of these vulnerabilities often involves multiple programming languages. 
Therefore, effective detection tools must be capable of handling cross-language code contexts, a point which is also supported by our findings in RQ3.

\begin{tcolorbox}[left=6pt,right=6pt,boxsep=-1mm]
At the package level, multi-language is a common characteristic in Python packages. Our analysis indicates that multi-lingual Python packages are potentially more susceptible to vulnerabilities.
At the commit level, the vulnerability samples within Python packages also involve various programming languages. However, only 1.6\% of the commits involve vulnerable functions in multiple languages, indicating that these vulnerabilities tend to be fixed within a single language.
\end{tcolorbox}
\begin{table}[tbp]
\footnotesize
\centering
\caption{Vulnerability types distribution of \benchname{}.}
\setlength{\tabcolsep}{2pt}
\label{tab:dataset_cwe}
\begin{tabular}{lccccc}
\toprule
\textbf{Type} & \textbf{\#Commits} &  \textbf{\#Functions}& \textbf{Avg. CVSS} \\  \midrule
Injection & 202 (17.5\%)& 411 (19.7\%)& 7.4\\
Improper Access Control & 133 (11.5\%)& 305 (14.6\%)& 7.3\\
Out-of-Bound Read/Write & 114 (9.9\%)& 174 (8.4\%)& 6.0\\
File Operation Error & 80 (6.9\%)& 165 (7.9\%)& 6.9\\
Improper Input Validation & 75 (6.5\%)& 109 (5.2\%)& 6.7\\
Calculation Error & 66 (5.7\%)& 79 (3.8\%)&4.6 \\
Sensitive Information Exposure & 60 (5.1\%)& 103 (4.9\%)&5.8\\
Request Forgery & 53 (4.6\%)& 109 (5.2\%)& 7.5 \\
Improper Resource Management & 54 (4.7\%)& 112 (5.4\%)& 6.5\\
NULL Pointer Dereference & 43 (3.7\%)& 53 (2.5\%)& 5.4 \\
Assertion Failures & 39 (3.4\%)& 46 (2.2\%)& 5.4 \\
Incorrect Synchronization & 38 (3.3\%)& 66 (3.2\%)& 5.6 \\
Redirect Error& 23 (2.0\%)& 47 (2.3\%)& 6.1\\
Use of Uninitialized Resource & 24 (2.1\%)& 29 (1.4\%)& 5.9\\
Improper Deserialization & 23 (2.0\%)& 39 (1.9\%)& 8.9\\
Incorrect Regular Expression & 22 (1.9\%)& 31 (1.5\%)& 6.1\\
Uncontrolled Recursion & 16 (1.4\%)& 24 (1.2\%)& 5.9\\
Improper Exception Handling & 16 (1.4\%)& 20 (1.0\%)& 5.3\\
Inefficient Algorithmic Complexity & 12 (1.0\%)& 39 (1.9\%)& 7.0\\
Incorrect Provision of Specified Functionality & 9 (0.8\%)& 27 (1.3\%)& 3.5\\
Incomplete Cleanup  & 6 (0.5\%)& 11 (0.5\%)& 7.2\\
Side Channel & 5 (0.4\%)& 8 (0.4\%)& 5.2\\
Others& 43 (3.7\%)& 70 (3.4\%)& 6.7\\
\midrule
\textbf{Total}& 1,157 & 2,082 & 6.5\\
\bottomrule
\end{tabular}
\end{table}

\subsubsection{Vulnerability Type Distribution}
Vulnerabilities come in many different types, each varying in detection difficulty. 
Beyond simply assessing whether a vulnerability detection method can find vulnerabilities, we are also interested in its performance when detecting different types of vulnerability. 
Therefore, 
we additionally annotate the \benchname{} dataset with CWEs from original vulnerability reports.
The 1,157 commit-level vulnerabilities in the \benchname{} dataset belong to 151 different CWE vulnerability types. 
We performed a simple clustering based on the mechanisms, the causes and the consequences of these CWE vulnerability types.
For example, CWE-125 (Out-of-bounds Read), CWE-787 (Out-of-bounds Write), CWE-120 (Buffer Copy without Checking Size of Input, 'Classic Buffer Overflow'), and CWE-122 (Heap-based Buffer Overflow) were merged into one category.
We list the details of the clustering in Appendix~\ref{sec:appendix}.

In Table~\ref{tab:dataset_cwe}, we list the distribution of vulnerability types in Python packages.
Injection vulnerabilities are the most common type, with 195 commits (394 functions), accounting for 17.5\% (19.7\%) of the total.
Injection vulnerabilities consist of 16 CWE vulnerability types, including SQL Injection, Command Injection, Parameter Injection, Cross-site Scripting (XSS) Injection, Static Code Injection, XML External Entity (XXE) Injection, CSV Formula Injection, and others.
Access control vulnerabilities are the second most common type, accounting for 11.5\% of the total commits (133 commits) and 14.6\% of the total functions (305 functions).
Access control vulnerabilities consist of 33 CWE vulnerability types, primarily including CWE-284 (Improper Access Control), CWE-287 (Improper Authentication), CWE-305/289/288/290/294 (Authentication Bypass by Primary Weakness/Alternate Name/Using an Alternate Path or Channel/Spoofing/Capture Replay), and CWE-304 (Missing Critical Step in Authentication), among others.
Following closely are vulnerability types such as Out-of-Bound Read/Write, File Operation Error, Improper Input Validation, and Calculation Error, which also occur relatively frequently.

From the vulnerability types we can spot a great diversity regarding their origins and attack scenarios. Vulnerabilities such as XSS Injection, Improper Access Control and Request Forgery are predominantly associated with web applications. On the other hand, vulnerabilities such as Out-of-Bound Read/Write, NULL Pointer Dereference and Use of Uninitialized Resource are typically linked to low-level C/C++ code. Additionally, 
Incorrect Synchronization relates to parallel execution.
This diversity in vulnerability type echoes Python's usage in different fields and may pose extra difficulty to automated static detectors, including both rule-based ones and ML-based ones. 

\begin{tcolorbox}[left=6pt,right=6pt,boxsep=-1mm]
\benchname{} exhibits a total of 151 distinct CWE vulnerability types. 
Among the vulnerability types, Injection, Improper Access Control, Out-of-Bound Read/Write are the most common types.
The vulnerability types within Python packages are diverse in their origins and attack scenarios, echoing Python's application in various fields and posing addition difficulty to automated static detectors.
\end{tcolorbox}

\subsection{RQ3: Evaluation of Rule-based Detectors}\label{subsec:eval_rule}

With a more accurately annotated vulnerability benchmark, we can perform a more precise evaluation of existing vulnerability detection methods.
Since rule-based and ML-based vulnerability detection methods typically operate at different levels (commit or function), we evaluate them separately. 
For rule-based vulnerability detection methods, we use the commit-level vulnerability dataset for evaluation.

Due to the limitations of rule-based static vulnerability detection methods, such as CodeQL, PySA, and Bandit, which can only detect vulnerabilities for a specific programming language at a time, 
we use samples in \benchname{} that involve only the Python language, accounting for 68.1\% (788/1,157) of the Commit vulnerability dataset.
In this \textit{PyVul (Python)} subset, we selected six vulnerability types with the highest occurrence frequency, totaling 244 commit-level vulnerability samples, to assess the efficiency of rule-based detectors, as shown in the second column of Table~\ref{tab:rule_based}.

We apply the detectors to scan the vulnerability samples from CWEs that they target and report the number of complete runs in the third column. Among the three detectors, PySA exhibits a notable number of failed runs. Out of 244 scans, only 54 of them finish within a one-hour time window without any interrupting runtime errors.

The detection results of rule-based static analysis methods are difficult to verify automatically. 
Even when the detection results identify the target vulnerability, the reported locations of the vulnerable code can vary from the patch, especially for taint-style vulnerabilities.
As a result, we manually interpret and verify each of the detection results that matches the CWE of the target-reported vulnerabilities.
The results of this manual verification are presented in the fifth column of Table~\ref{tab:rule_based}.
Overall, the best-performing detector in our evaluation, CodeQL, successfully detects 10.8\% (23/212) of these reported real-world vulnerabilities. Bandit detects 5.3\% (10/189) of these vulnerabilities, and PySA fails to detect any of them.
For the detection of specific CWE categories of vulnerability, CodeQL again demonstrates the best performance and detects 30.0\% (15/50) of the CWE-22 vulnerabilities.

We further list the total number of warnings given by the detectors after scanning the target vulnerabilities in the fourth column of Table~\ref{tab:rule_based}.
In total, PySA generates 168 warnings for 54 vulnerability samples, CodeQL generates 6,078 warnings for 212 samples, and Bandit generates 323,023 warnings for 189 samples. Notably, on average, Bandit outputs 1,709 warnings for each sample.
The high volume of warnings produced by Bandit underscores a potentially significant, or even impractical, manual auditing effort.

Overall, none of the evaluated detectors are capable of effectively identifying vulnerabilities in \textit{PyVul (Python)}. In addition, low completion rate and excessive warning numbers further undermine the applicability of the detectors in real-world scenarios.



\begin{table}[tbp]
\centering
\caption{Performance of rule-based detectors.}
\vspace{-0.5em}
\setlength{\tabcolsep}{5pt}
\footnotesize
\begin{threeparttable} 
\begin{tabular}{crrrrr}
\toprule
\textbf{Detector}                & \textbf{CWE}     & \textbf{\#Commit} & \textbf{\#Complete}&\textbf{\#Warnings}& \textbf{\makecell{\#Verified \\Positives}} \\
\midrule
\multirow{7}{*}{CodeQL} & CWE-79  & 73      &73   & 1,633   & 5 \\
                        & CWE-22  & 50    &50    & 1,414   & 15 \\
                        & CWE-400 & 37    &37     &    1,302   & 2\\
                        & CWE-362 &     32   & -  &   -    & -\\
                        & CWE-89 &    29   &29    &     1,341    & 1\\
                        & CWE-352 &    23   &23    &     388   & 0\\
\cmidrule{2-6}
                        & Total   & 244   &  212  &   6,078 & 23 \\
\midrule
\multirow{7}{*}{PySA}   & CWE-79  & 73    &24    & 25     & 0 \\
                        & CWE-22  & 50     &30   & 143    & 0 \\
                        & CWE-400 &   37    &-    &  -     & -\\
                        & CWE-362 &    32   &-    &   -    & -\\
                        & CWE-89 &     29   &0   &  0     & 0\\
                        & CWE-352 &    23  &-     &  -     &- \\
\cmidrule{2-6}
                        & Total   & 244   &54   &   168  & 0 \\
\midrule
\multirow{7}{*}{Bandit} & CWE-79  & 73   &73     & 73,933 & 8 \\
                        & CWE-22  & 50    &50    & 127,273     & 0 \\
                        & CWE-400 &     37   &37   &     90,889  & 0\\
                        & CWE-362 &     32   &-    &  -     & -\\
                        & CWE-89 &     29   &29   &  30,928     & 2\\
                        & CWE-352 &  23      &-   &  -     & -\\
\cmidrule{2-6}
                        & Total   & 244    &189   & 323,023 & 10 \\
\bottomrule
\end{tabular}
\begin{tablenotes}
\item CWE-79: Cross-site Scripting; CWE-22: Path Traversal; CWE-400: Uncontrolled Resource Consumption; CWE-362: Race Condition; CWE-89: SQL Injection; CWE-352: Cross-Site Request Forgery.
\end{tablenotes}
\end{threeparttable}
\label{tab:rule_based}
\end{table}

\begin{tcolorbox}[left=6pt,right=6pt,boxsep=-1mm]
The best-performing rule-based vulnerability detection approach detects merely 10.8\% of the real-world vulnerabilities in \textit{PyVul (Python)}. Additionally, two issues spotted in the evaluated detectors, low completion rate and excessive volume of warnings, potentially undermine their applicability in the real world.
\end{tcolorbox}

\subsubsection{Limitations of Rule-based Detectors}
To understand the underlying reasons for the inefficiency of these static analysis tools, we randomly select and manually audit 30 cases where detection failed and summarize the causes of detection failures for each of the six CWE vulnerability types.
The following lists the discussion of three CWEs and we present the rest in Appendix~\ref{cwe_review}.

\noindent\textbf{CWE-79: Cross-site Scripting.}
Cross-site Scripting (XSS) vulnerabilities are a type of injection vulnerability where user-supplied data gets rendered in web pages without adequate safety checks, resulting in potential malicious code execution. The examined XSS vulnerabilities can be categorized into three types:
1) Reflected XSS \textbf{(12/30)}.
2) Stored XSS \textbf{(16/30)}.
3) Improper URL parameter validation leading to potential XSS in downstream applications \textbf{(2/30)}.
Reflected XSS and stored XSS, the predominant types of XSS identified in Python packages, ultimately share the same taint sources, i.e., user input from remote flows, and taint sinks, i.e., server responses.
Involved in web applications, these vulnerabilities exhibit complex data flows and require sophisticated taint analysis, which is evidenced by the variety of the fixing locations. 
For reflected XSS, we observe two commonly adopted fixes in real-world reports:
1) sanitizing the fields of web pages (7/12),
2) sanitizing the data parsed from user requests (3/12).
As evidenced by the fixes, accurate detection of reflected XSS requires cross-language taint analysis that takes both server-side code and client-side code into consideration.
Neither CodeQL or PySA supports cross-language taint analysis.

Stored XSS extends further. We observe fixes have been applied in various locations, including: 1) proper sanitization in object-\linebreak relational mappings (ORM) (4/17),
2) proper sanitization in response crafting (5/17),
3) proper sanitization in client-side code (2/17),
4) setting Content Security Policy (CSP) in the server's configuration (4/17).
This requires a more comprehensive scope of analysis. 
In addition, stored XSS is considered a high-order taint-style vulnerability~\cite{zhang2021statically} where there exist two phases of taint flows~\cite{su2023splendor}. In the first phase, taint flows from remote flow sources to data storage sinks. In the second phase, the exact taint is loaded from data storage and flows to server response generation sinks. As such, for detecting stored XSS, the detectors are required to identify the stored location of the taint and bridge the two phases. 
Neither of the two detectors makes any effort to model these complex taint flows for stored XSS.

The third type of XSS, improper validation of URL parameters leading to potential XSS in downstream applications, commonly occurs in non-standalone packages—those that are not self-contained and primarily serve as utility providers for other applications. 
Effective taint analysis relies on the precise definition of taint sources and sinks, which can be predefined by the detectors or supplied by the detector users. Both CodeQL and PySA have a comprehensive range of predefined sources and sinks. However, for non-standalone packages where their downstream applications need to be taken into consideration, these predefined sources and sinks can hardly be effective.
For example, in the context of a web framework such as Django~\cite{Django}, a function parameter may be used by downstream applications to pass user-supplied data, thus qualifying as a taint source.
Vulnerabilities associated with these package-specific sources and sinks cannot be automatically identified by the subject detectors.

Apart from taint analysis rules targeting specific types of XSS, Bandit, and CodeQL have rules checking HTML escaping project-wide to address general XSS, such as checking if the Jinja2 environment is set to auto-escape.
Such rules mitigate certain XSS vulnerabilities but face several problems: 1) Jinja2, despite its popularity, is not universally adopted across all Python web applications. Even for those that employ Jinja2 templates, it is not guaranteed that all client-side web pages are generated using Jinja2.
2) Dynamic content on web pages that uses JavaScript can also introduce XSS vulnerabilities;
3) Setting Jinja2 environment to auto-escape may be against the project's business logic.
For example, the project \emph{nbdime} involved with CVE-2021-41134~\cite{cve202141134} is a tool for diffing and merging of Jupyter notebooks and offers web-based extensions.
In such a case where user-uploaded code is displayed on web pages, project-wide automatic escaping may not be a viable solution.

\noindent\textbf{CWE-22: Path Traversal.}
Path traversal refers to a situation where an application receives unvalidated user input as parameters for file-related operations, such as reading or viewing files. 
These parameters contain special characters (e.g., `..` and `/`) that can be used to bypass protection mechanisms, gain unauthorized access to protected files or directories, or overwrite sensitive data.
Several types of path traversal vulnerability are spotted: 1) Improper use of other packages \textbf{(4/30)}; 2) Unawareness of behavioral differences of used APIs from other packages when executed on different operating systems \textbf{(3/30)}; 3) Missing validation in certain taint paths \textbf{(22/30)}.
The first two causes are not considered by any of these evaluated detectors. 
For the static detection of the third cause, taint analysis is the typical approach.
CodeQL and PySA support static taint analysis. 
However, 
four factors hinder their performance in the detection of CWE-22:
\begin{itemize}[leftmargin=*]
\item Lack of package-specific taint specifications. 
Non-standalone packages require package-specific taint specifications.
\item Lack of accurate type information. This is an inherited challenge for Python static analyzers. 
As a dynamic language, variable types in Python are determined at run time.
Without type information, static modeling of data flows can be largely incomplete, substantially limiting the effectiveness of taint analysis built upon it. 
Implementing type inference can mitigate this challenge.
However, neither of the subject detectors incorporate any form of type inference.
\item Limited handling of Python's complex language features. Python’s advanced language features, such as higher-order functions and dynamic features~\cite{Yang2022complexpython}, frequently present in the examined packages. Incomplete addressing of these features further contributes to incomplete data flow modeling.
\item Complex data flows in web applications. Web applications are frequently spotted in CWE-22 reports. 
The inherent complexity of web applications arises from their interaction with client-side components and their capability to execute multiple routes concurrently, often resulting in intricate data flows.
Neither of the detectors effectively models these intricate data flows.
\end{itemize}

\noindent\textbf{CWE-352: Cross-Site Request Forgery (CSRF).}
CSRF is an attack that allows the attacker to exploit a user's authentication credentials on a logged-in website and send malicious requests to that site. 
Two causes of CSRF are identified in the examined reports:
1) Using GET requests to modify database \textbf{(16/23)}. GET requests are supposed to be used only for viewing data, and using GET requests to change anything in the database is not protected by any CSRF protection policies;
2) CSRF protection is not applied to certain pages \textbf{(6/23)} (e.g., not setting CSRF tokens for certain forms).

Among the three evaluated detectors, only CodeQL includes a rule targeting CSRF vulnerabilities. 
However, CodeQL over-simplifies the vulnerability without addressing vulnerable code patterns in the real world. 
CodeQL operates under the assumption that modern web frameworks include built-in CSRF protections and that vulnerabilities only arise when these protections are explicitly disabled. 
CodeQL's rule checks if a web framework is used and whether CSRF protections have been disabled in the global settings.
Yet, we have not observed such cases in real-world vulnerability reports. 
Most CSRF vulnerabilities exist with certain CSRF protection turned on. As such, CodeQL's rule is completely ineffective against these real-world vulnerabilities.



Our evaluation has yielded two significant insights regarding the current state of rule-based detectors:
\begin{itemize}[leftmargin=*,before=\vspace{-1mm},after=\vspace{-1mm}]
\item In terms of the detector rules, we observe a significant discrepancy between the assumptions of the evaluated detectors and the real-world security landscape. The evaluated rule-based detectors tend to oversimplify the vulnerabilities, resulting in either a high volume of false positives or complete missing detection of real-world cases.
\item In terms of the detector architecture, current rule-based vulnerability detectors for Python lack: 1) support for high-order vulnerabilities, web application-related vulnerabilities can be high-order, such as stored XSS; 2) capability of modeling cross-language vulnerabilities; 3) accurate data flow modeling that addresses Python's language features.
\end{itemize}

\begin{tcolorbox}[left=6pt,right=6pt,boxsep=-1mm]

Our empirical evaluation reveals two primary limitations in the current state of rule-based vulnerability detectors for Python: 1. significant discrepancy between the assumptions of the detectors and real-world security scenarios, 2. lack of support for high-order and cross-language vulnerabilities, and Python's language features.
\end{tcolorbox}

\subsection{RQ4: Evaluation of ML-based Detectors}
In this RQ, we examine how well the real-world Python package vulnerabilities in \benchname{} can be identified by ML-based detectors. 
Existing approaches~\cite{chen2023diversevul, ding2024vulnerability} commonly detect vulnerabilities at the function level. To address this requirement, we utilize function-level samples from \benchname{} for our evaluation.


\setlength{\textfloatsep}{3pt}
\begin{table}[tbp]
\centering
\footnotesize
\caption{Performance of ML-based approaches in detecting vulnerabilities in \benchname{}. }
\vspace{-0.5em}
\label{tab:binary}
\setlength{\tabcolsep}{2pt}
\begin{tabular}{@{}lccccccccc@{}}
\toprule
\textbf{Model} & \textbf{Train} & \textbf{Test} & \textbf{Dataset} & \textbf{Invalid} & \textbf{Acc.} & \textbf{Prec.} & \textbf{Recall} & \textbf{F1}  &\\
\midrule
\multirow{ 2}{*}{CodeQwen1.5-Chat}  & \multirow{ 2}{*}{-} & \multirow{ 2}{*}{300} &Paired&  19 & 53.4\% & 51.9\% & 59.1\% & 55.3\%  \\\cline{4-10} 
& & &Non-paired&23& 58.5\%& 57.4\% & 66.9\%& 61.8\% \\
\hline
\multirow{ 2}{*}{GPT-3.5 Turbo}  & \multirow{ 2}{*}{-} & \multirow{ 2}{*}{300} &Paired& 0 & 49.5\% & 49.5\% & 60.4\% & 54.4\%\\ \cline{4-10}
& & &Non-paired& 3 & 50.8\% & 50.8\% & 61.1\% & 55.5\% \\
\hline
\multirow{ 2}{*}{GPT-4}  & \multirow{ 2}{*}{-} & \multirow{ 2}{*}{300} &Paired& 0 & 58.0\% & 52.1\% & 32.7\% & 40.2\%\\ \cline{4-10}
& & &Non-paired& 0 & 51.3\% & 65.8\% & 33.3\% & 44.3\% \\
\hline
\multirow{ 2}{*}{\begin{tabular}{@{}l@{}}CodeQwen1.5-Chat \\finetuned\end{tabular}}  & \multirow{ 2}{*}{300} & \multirow{ 2}{*}{300} &Paired& 0 & 51.0\% & 50.7\% & 72.0\% & 59.5\%  \\\cline{4-10} 
& & &Non-paired& 0&62.7\% & 60.1\% & 75.3\% & 66.9\%    \\
\hline
\multirow{ 2}{*}{\begin{tabular}{@{}l@{}}GPT-3.5 Turbo \\finetuned\end{tabular}}  & \multirow{ 2}{*}{300} & \multirow{ 2}{*}{300} &Paired&  0 & 50.0\% & - & 0\% & 0\% \\\cline{4-10}
& & &Non-paired& 0 & 67.7\% & 63.9\% & 81.3\% & 71.6\% \\
\hline
\multirow{ 2}{*}{\begin{tabular}{@{}l@{}}GPT-3.5 Turbo \\finetuned\end{tabular}}  & \multirow{ 2}{*}{1500} & \multirow{ 2}{*}{300} &Paired& 0 & 50.0\% & 50.0\% & 99.3\% & 66.5\% \\\cline{4-10}
& & &Non-paired& - & - & - & - & - \\

\bottomrule
\end{tabular}
\end{table}
\setlength{\floatsep}{6pt}

\subsubsection{Performance of Vulnerability Prediction}\label{subsec:paired_data}


To train and evaluate the ML models, we require both vulnerable and non-vulnerable samples. 
There are several strategies for curating non-vulnerable samples. 
One approach is to collect the patched versions of vulnerable functions, while another is to gather unrelated functions.
Using the patched versions creates a more challenging scenario, as vulnerable and benign samples tend to be very similar, differing by only a few lines of code. 
This requires ML-based approaches to have a deeper understanding of the intrinsic characteristics of the vulnerabilities.
In our experimental setup, we use both strategies, referring to them as paired and non-paired datasets. 
To create the non-paired dataset, we compile a pool of benign samples from two sources: 1) 
newly added functions in \benchname{}'s commit, which do not have pre-fix versions,
and 2) patched versions of functions that are labeled by \textit{LLM-VDC} as irrelevant to the vulnerabilities.
This results in a total of 1,462 functions. 
For each vulnerable sample, we randomly select a benign sample written in the same programming language from this pool.

The three LLMs introduced in Section~\ref{sec:ml-approaches} are evaluated under two different setups: direct prompting and fine-tuning. 
For the evaluation across all settings, we use a total of 300 samples, which include 150 vulnerable samples and 150 benign samples, following~\cite{purba2023software}.
In the direct prompting setup, we adopt a zero-shot approach and follows~\cite{ding2024vulnerability} for the chain-of-thought prompt. 
To the best of our knowledge, ~\cite{ding2024vulnerability} represents one of the most recent studies evaluating the capability of LLMs to detect vulnerabilities using direct prompting.
Furthermore, we fine-tune CodeQwen1.5-Chat and GPT-3.5 Turbo using a different set of 300 samples, following the settings described in Section~\ref{sec:exp-setup}. 
To examine the effect of data volume on fine-tuning, we also conduct an additional experiment where we fine-tune GPT-3.5 Turbo with a larger dataset of 1,500 samples. 
For all experimental setups, we evaluate the models' performance using various metrics and present the results in Table~\ref{tab:binary}.
As LLMs occasionally generate answers that are not in the required format, we report the number of such invalid answers in the fifth column.

As shown in Table~\ref{tab:binary}, LLMs without additional adjustments achieve accuracies ranging from 49.5\% to 58.5\% and F1 scors ranging from 40.2\% to 61.8\%, which are only marginally better than random guesses, indicating that they do not inherently support vulnerability detection tasks.
Fine-tuning on non-paired data greatly improves the performance of LLMs in terms of vulnerability detection, making it a potentially promising direction. 
Specifically, GPT-3.5 Turbo and CodeQwen1.5-Chat fine-tuned on 300 non-paired data yields a 29.0\% and 8.3\% increase in F1 scores respectively, achieving F1 scores of 71.6\% and 66.9\%.

However, fine-tuning on paired data reveals a severe problem. 
LLMs fine-tuned on paired data achieve even worse performance than in a zero-shot setting. 
Specifically, CodeQwen1.5-Chat, despite an increase of 4.2\% in F1 score, shows a decrease of 2.4\% in accuracy. GPT-3.5 Turbo completely fails to derive any meaningful learning from the paired data and consistently predicts every test case as vulnerable.
Moreover, despite being trained on a larger dataset of 1500 samples, GPT-3.5 Turbo does not demonstrate any learning improvements, maintaining an accuracy of 50.0\%.
This indicates that LLMs are not able to differentiate the vulnerable functions and their patched version. In real world, vulnerable code and benign code are largely similar and often differ in a small number of lines.
The inability of LLMs to differentiate between these subtle variations suggests that current LLM-based approaches may struggle to provide practical utility in real-world applications.


\begin{tcolorbox}[left=6pt,right=6pt,boxsep=-1mm]

The LLM-based vulnerability detection approaches, though achieve relatively promising performance on non-paired data with a best F1 score of 75\%, fail to differentiate vulnerable sample with their largely similar patched version, indicating their limited capability in real-world scenarios. 

\end{tcolorbox}

\setlength{\floatsep}{6pt}
\begin{table}[t]
\centering
\footnotesize
\caption{Performance of GPT-3.5 Turbo when fine-tuned and tested on different CWE types of vulnerabilities.
}
\vspace{-0.5em}
\label{tab:performance_by_cwe}
\setlength{\tabcolsep}{2pt}
\begin{threeparttable} 
\begin{tabular}{lccccccc}
\toprule
\textbf{CWE} & \textbf{Train} & \textbf{Test} & \textbf{Invalid} & \textbf{Acc.} & \textbf{Prec.} & \textbf{Recall}& \textbf{F1}  \\
\midrule
\begin{tabular}{@{}l@{}}CWE-79 \\(Cross-site Scripting)\end{tabular}
 & 232 & 58 & 0 & 58.1\% & 56.4\% & 71.0\% &62.9\% \\\hline
\begin{tabular}{@{}l@{}}CWE-22 \\(Path Traversal)\end{tabular}
 & 160 & 42 &  0 & 72.5\% & 90.9\% & 50.0\%& 64.5\% \\\hline
\begin{tabular}{@{}l@{}}CWE-20 \\(Improper Input Validation)\end{tabular}
&  158 & 40 & 0 & 75.0\% & 69.2\% & 90.0\%& 78.3\% \\\hline
\begin{tabular}{@{}l@{}}CWE-476 \\(NULL Pointer Dereference)\end{tabular}
& 84 & 22 & 0 & 68.2\% & 62.5\% & 90.9\%& 74.1\%\\\hline
\begin{tabular}{@{}l@{}}CWE-125 \\(Out-of-bound Read)\end{tabular}
 & 84 & 22  & 0& 77.3\% &  75.0\% & 81.8\% & 78.3\%\\
\bottomrule
\end{tabular}
\end{threeparttable}
\end{table}

\subsubsection{Performance Discrepancies Across CWEs}
We further investigate the performance of ML-based approaches when they are fine-tuned and tested on different types of CWE vulnerabilities. 
Due to the limited amount of data available for each CWE category, we select the five most prevalent CWEs from our function-level benchmark and utilize all vulnerable samples from each category.
We adopt a non-paired setting because LLMs struggle to learn effectively from paired data, which leads to uninformative metrics. 
For this evaluation, we choose GPT-3.5 Turbo, as it shows the highest F1 score when fine-tuned on non-paired data in Section~\ref{subsec:paired_data}.
We follow the same method as Section~\ref{subsec:paired_data} to prepare the non-paired datasets.
The datasets curated for each selected CWE category are then divided using an 80/20 split for fine-tuning and testing.


The data in Table~\ref{tab:performance_by_cwe} indicates that fine-tuning GPT-3.5 Turbo on different CWE categories results in varying performance levels. Notably, the performance on CWE-79 and CWE-22 is significantly lower compared to other categories. Specifically, CWE-79, despite being trained on the largest sample size of 232, attained the lowest F1 score of 62.9\%. In contrast, CWE-125, which was trained on a smaller sample set of 84, achieved a higher F1 score of 78.3\%.




We further investigate the performance discrepancy between CWE-79 and CWE-125 by manually examining 50 
vulnerable functions from each category. In comparing CWE-79 with CWE-125, we identify two factors that may contribute to the poor performance associated with fine-tuning CWE-79: the specific characteristics of the CWE category and the training method used.

1) \textbf{The great variance of the vulnerable functions.}
CWE-79 (Cross Site Scripting) is a taint-style vulnerability~\cite{yamaguchi2015automatic}. The taint flows of CWE-79 are typically complex and involve multiple functions.
To fix the vulnerability, the sanitizations can be applied at different places of the taint flows.
In real-world instances, we have observed sanitization being applied to diverse locations—from data storage functions interacting with databases to various segments of client-side code like HTML, JavaScript, or template creation, to even within server configurations.
In this case, in the function-level setting where the changed functions prior to the fixing commit are marked as vulnerable, a diverse range of vulnerable function samples may be observed.
Such great variance can confuse the model and hinder its ability to learn effective patterns for identifying vulnerabilities.




\definecolor{addcolor}{RGB}{218,251,225}
\definecolor{delcolor}{RGB}{255,235,233}
\definecolor{ngreen}{RGB}{16,113,15}
\definecolor{nred}{RGB}{159,0,22}

\lstdefinelanguage{JavaScript}{
  keywords={break, case, catch, continue, debugger, default, delete, do, else, false, finally, for, function, if, in, instanceof, new, null, return, switch, this, throw, true, try, typeof, var, void, while, with},
  morecomment=[l]{//},
  morecomment=[s]{/*}{*/},
  morecomment=[f][\color{ngreen}]{+},
morecomment=[f][\color{nred}]{-},
  morestring=[b]',
  morestring=[b]",
  moredelim=[s][\color{codepurple}]{/}{/g},
  ndkeywords={class, export, boolean, throw, implements, import, this},
  sensitive=true
}

\newcommand\realnumberstyle[1]{}

\makeatletter
\newcommand{\zebra}[1]{%
    \ifnum\value{lstnumber}=1\relax
        \color{addcolor}%
    \else\ifnum\value{lstnumber}=2\relax
        \color{addcolor}%
    \else\ifnum\value{lstnumber}=3\relax
        \color{addcolor}%
    \else\ifnum\value{lstnumber}=11\relax
        \color{delcolor}%
    \else\ifnum\value{lstnumber}=12\relax
        \color{addcolor}%
    \else
        \color{white} 
    \fi\fi\fi\fi\fi
    \rlap{\hspace*{\lst@numbersep}%
    \color@block{\linewidth}{\ht\strutbox}{\dp\strutbox}%
    }%
}
\makeatother

\setlength{\floatsep}{8pt}
\begin{figure}
    \centering
    \begin{lstlisting}[language=JavaScript, style=mystyle,numbers=none,frame=single,rulecolor=\color{black}
   ]
+ function htmlEntities(str) {
+     return String(str).replace(/&/g, '&amp;').replace(/</g, '&lt;').replace(/>/g, '&gt;').replace(/"/g, '&quot;');
+ }
  ...
  {
      data: "ticket",
      render: function (data, type, row, meta) {
          if (type === 'display') {
              data = '<div class="tickettitle"><a href="' + get_url(row) + '" >' +
                  row.id + '. ' +
-                 row.title + '</a></div>';
+                 htmlEntities(row.title) + '</a></div>';
          }
          return data
      }
  }\end{lstlisting}
    \vspace{-1em}
    \caption{Example of XSS vulnerability CVE-2021-3945~\cite{cve20213945}. The fix in commit 2c7065e~\cite{commit2c7065e}.}
    \label{fig:rq4_example_1}
\end{figure}

Figure~\ref{fig:rq4_example_1} shows an example of XSS vulnerability.
This vulnerability arises when a user creates a ticket with a malicious title that injects html code. As the ticket titles are never sanitized, when the tickets are rendered on the admin's page, the injected HTML code will take effect.
The fix adopted by the developers sanitizes the ticket titles in the \verb|render| function where the tickets are rendered.
Alternatively, the vulnerability could also be effectively mitigated by sanitizing the data before it is stored in the database, specifically within the \verb|form_valid| function, as shown in Figure~\ref{fig:rq4_example_2}.
In this instance, if developers opt to sanitize the data in \verb|render|, then \verb|render| is identified as the vulnerable function; if sanitization occurs in \verb|form_valid|, then \verb|form_valid| is marked as vulnerable.
In contrast, the patches of CWE-125 (Out-of-bound Read) typically exhibit shorter taint flows. When the vulnerabilities are processed as vulnerable functions, the models are able to observe more stable vulnerable code patterns compared to CWE-79.



2) \textbf{Inability to see important context.}
In the patches of CWE-79 vulnerabilities, sanitizations are commonly implemented as new functions and then applied to the input data.
However, in the function-level setting, the model may not be able to access the content of such sanitization functions, resulting in additional difficulty in differentiating between the vulnerable code and fixed code pair.

For instance, in the same example of Figure~\ref{fig:rq4_example_1}, the developers create a sanitization function \verb|htmlEntities|, which sanitizes potential HTML injection, and invoke it in the \verb|render| function. 
In the function-level settings, including the ones that take a step further and group functions implicated in cross-function vulnerabilities, such as~\cite{sejfia2024toward}, as the function \verb|htmlEntities| is newly created and does not have a pre-existing vulnerable counterpart, it will not be included in the dataset that is used to train or test models.
Consequently, this omission impedes the model's ability to discern between the pre-fix and post-fix versions of the \verb|render| function.

Therefore, we argue that the typical function-level setting is problematic to real-world scenarios, failing to capture the characteristics of real-world vulnerabilities.
This echoes the observation reported by Risse et al.~\cite{risse2024top}.
Future work is encouraged to explore innovative training methods to incorporate relevant contexts of vulnerabilities and enable models to effectively learn vulnerable code patterns despite the variance of fixing patches.

\begin{tcolorbox}[left=6pt,right=6pt,boxsep=-1mm]
The performance of fine-tuned LLMs for vulnerability detection varies significantly across different CWE categories of vulnerabilities.
The commonly adopted function-level setting fails on complex real-world vulnerabilities for two reasons: 1) the great variance of the vulnerable functions, and 2) the potential absence of important context.
\end{tcolorbox}
\begin{figure}
    \centering
\begin{lstlisting}[language=Python,numbers=none,frame=single,]
def form_valid(self, form):
    request = self.request
    if text_is_spam(form.cleaned_data['body'], request):
        return render(request, template_name='helpdesk/public_spam.html')
    else:
        ticket = form.save(user=self.request.user if self.request.user.is_authenticated else None)
        try:
            return HttpResponseRedirect('%s?ticket=%s&email=%s&key=%s' % (
                reverse('helpdesk:public_view'),
                ticket.ticket_for_url,
                urlquote(ticket.submitter_email),
                ticket.secret_key)
                                        )
        except ValueError:
            return HttpResponseRedirect(reverse('helpdesk:home'))
\end{lstlisting}
    \vspace{-1em}
    \caption{Relevant data storing function of XSS vulnerability CVE-2021-3945. }
    \label{fig:rq4_example_2}
\end{figure}



\section{Related Work}

\textbf{Vulnerability empirical study.}
Many empirical studies~\cite{jimenez2016empirical,tan2014bug,linares2017empirical} have been conducted to study vulnerabilities in other software systems or software ecosystems.
Tan et al.~\cite{tan2014bug} analyzed around 2k real-world vulnerabilities in Linux kernel, Mozilla, and Apache and yielded several guidelines for developing corresponding detectors. 
Linares et al.~\cite{linares2017empirical} conducted a large-scale empirical study to characterize different types of vulnerability that affect Android apps.
Regarding the Python ecosystem, Alfadel et al.~\cite{alfadel2023empirical} make the first move to study the propagation and life span of Python security vulnerabilities. Besides, there has been research to study the bugs in machine learning (ML) libraries in Python~\cite{harzevili2023characterizing,zhang2018empirical,thung2012empirical,jia2021symptoms}.
Despite these efforts, the characteristics of security vulnerabilities in the whole Python package ecosystem have not been well studied, and how well current vulnerability detection tools perform on real-world vulnerabilities in Python packages remains unknown.

\textbf{Vulnerability datasets.}
Different datasets~\cite{fan2020ac,chakraborty2021deep,nikitopoulos2021crossvul,wartschinski2022vudenc,chen2023diversevul,ding2024vulnerability} have been presented to facilitate vulnerability detection.
Apart from the datasets we have compared \benchname{} with, notably, the ReVeal~\cite{chakraborty2021deep} dataset was labeled using the patches to known security issues at Chromium security issues and Debian security tracker.
BigVul~\cite{fan2020ac}
collect vulnerability-fixing commits from CVE records in the NVD. 
Cheng et al.~\cite{chen2023diversevul} presented DiverseVul with their empirical study, a new C/C++ vulnerable source code dataset that is 60\% larger than the previous largest dataset for C/C++, and the most diverse compared to all previous datasets. 
In contrast, our dataset is collected with a focus on vulnerabilities in Python packages and cleansed with the assistance of LLMs to achieve high label accuracy.

\textbf{Vulnerability dataset cleansing.}
Apart from PrimeVul~\cite{ding2024vulnerability}, ReposVul~\cite{wang2024reposvul} also targets the inaccurate labels in vulnerability datasets by combining LLMs and static vulnerability detectors. 
ReposVul determines a file as related to a vulnerability fix if both LLMs indicate its relevance and static vulnerability detectors identify vulnerabilities in its before-fixing version.
Comparing to ReposVul, our cleansing method, LLM-VDC, cleanses the datasets at a finer granularity with a higher accuracy. In addition, since ReposVul inherently depends on the outputs of static vulnerability detectors, its resulting dataset is unsuitable as a benchmark for evaluating static detectors.


\textbf{Vulnerability detection.}
Rule-based static vulnerability detection has been a commonly adopted approach, characterized by scalability and comprehensiveness. Previously, the method has been extensively explored in the context of statically typed languages~\cite{arzt2014flowdroid,yamaguchi2014modeling,li2015iccta,wei2018amandroid,zhang2021statically}. 
Recently, its application in dynamic languages such as JavaScript has also become a popular research area~\cite{kashyap2014jsai,khodayari2021jaw,li2022mining,kang2023scaling}.
However, rule-based static vulnerability detection in Python is yet to be explored by the research community, which may be attributed to Python's complex features~\cite{Yang2022complexpython}.
Even the most fundamental elements of static analysis, such as call graphs, have only recently been explored~\cite{salis2021pycg}.
The state-of-the-art static vulnerability detectors in Python are predominantly developed by industry, with notable tools such as CodeQL~\cite{CodeQL}, PySA~\cite{PySA}, and Bandit~\cite{Bandit} leading the efforts.

For ML-based vulnerability detection, previous papers have used LSTM~\cite{li2018vuldeepecker,wartschinski2022vudenc}, CNNs and RNNs~\cite{russell2018automated}, Bidirectional RNNS~\cite{li2021sysevr}, and Graph Neural Networks~\cite{chakraborty2021deep,mirsky2023vulchecker,zhou2019devign} to detect vulnerable source code. Among them, only VUDENC~\cite{wartschinski2022vudenc} was trained and evaluated on Python code. VUDENC applies a word2vec model to identify semantically similar code tokens and to provide a vector representation, and then uses an LSTM network to classify vulnerable code token sequences.
\section{Discussion \& Threats to Validity}
\textbf{Discussion.} 
Our LLM-assisted cleansing method labeled 72 out of 8,374 vulnerable functions as "4) no decision can be made". 
To ensure the integrity and validity of the evaluation on this automated cleansing method, especially when measuring the precision and recall,
we exclude the commits associated with these 72 functions from our dataset. 
Future work could explore altering the composition of contextual information provided to LLMs or incorporating additional context to help LLMs resolve such cases.

In our empirical evaluation of vulnerability detectors, we evaluated current rule-based and ML-based detectors and investigated their limitations independently. A direct comparison between these two methodologies was not conducted due to inherent differences in their operational granularities. Rule-based detectors scan whole projects and locate vulnerabilities precisely with detailed information such as the causes of the vulnerabilities and taint flow paths, while current ML-based detectors typically analyze individual functions and solely classify them as vulnerable or not.

\textbf{Threats to validity. }For RQ1 and RQ3, the dataset labels and the rule-based detectors' results are validated manually. 
The reliability of these decisions can be influenced by factors such as the evaluators' expertise in relevant areas and their personal interpretations of the vulnerabilities. 
To mitigate potential biases, we involve two authors, both with solid backgrounds in Python programming and software security, to independently assess the correctness of the labels and then resolve disputes. 
We additionally measure their agreement level with Cohan’s Kappa before any consensus has been reached, which is 0.718 for RQ1 and 0.601 for RQ3 and within the range of fair to good.

In all experiments in this study where humans are involved, there exist cases where the participant cannot make the decisions either because of limited descriptions or erroneously attached commits in the vulnerability reports. As the number of such cases is relatively small, we expect they do not affect the overall conclusions. For example, in RQ3, there is one reviewed path traversal vulnerability report that we cannot decide the type and the cause of it, which we expect does not affect our overall observations regarding this category of vulnerabilities.

\section{Conclusion}
In this paper, we presented \benchname{}, the first large-scale, high-quality benchmark suite of Python-package vulnerabilities, consisting of 1,157 vulnerable repository snapshots, and 2,082 vulnerable functions along with their respective patched versions.
We introduce LLM-VDC, an LLM-assisted data cleansing method, to cleanse \benchname{} and achieve a 62.1\% to 74.1\% improvement in function-level label accuracy compared to previous automatically collected vulnerability datasets, underscoring the effectiveness of our cleansing method in vulnerability labels.
Utilizing \benchname{}, we further evaluate current rule-based and ML-based static vulnerability detectors in Python. Our experimental results reveal that none of the current approaches is satisfactory for detecting these real-world vulnerabilities in Python packages.
Additionally, our empirical study delves into the limitations of these detectors, offering critical insights to fuel future development of static vulnerability detectors.

\bibliographystyle{ACM-Reference-Format}
\bibliography{main}
\appendix
\section{Vulnerability Span Analysis}
\begin{figure}[tbp]
    \centering
    \includegraphics[width=\linewidth]{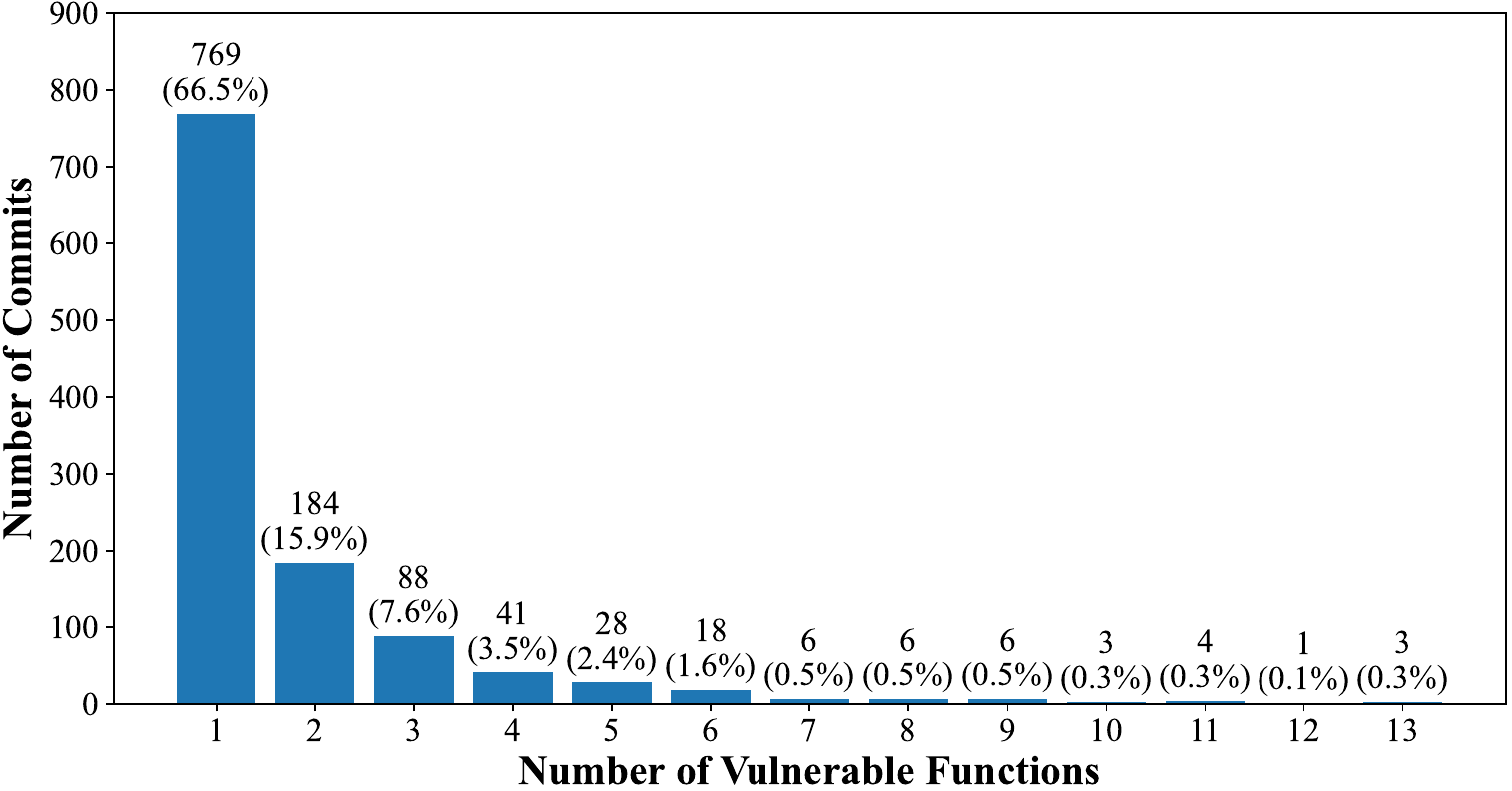}
    \caption{Vulnerable functions count distribution of \benchname{}.}
    \label{fig:cross_function_distribution}
\end{figure}

Span analysis aims to examine the number of functions related to a vulnerability. 
It provides crucial insights for detecting and addressing vulnerabilities, as the span reveals the minimum context required for effective analysis. 
However, a precise measurement of the span has not yet been obtained due to the inaccurate identification of code changes relevant to vulnerabilities in prior benchmarks. Given the high quality of \benchname{}, we evaluate how many functions are involved in the vulnerabilities of Python packages and present the statistics in Figure~\ref{fig:cross_function_distribution}.
The number of functions involved in the vulnerabilities of \benchname{} ranges from 1 to 13, with fewer vulnerabilities observed in more expansive categories. On average, each vulnerability in \benchname{} is linked to 1.8 function. 
In particular, 503 (43.5\%) vulnerabilities involve more than one vulnerable function. 
These cross-function vulnerabilities are associated with an average of 2.6 vulnerable functions. 
The prevalence of cross-function vulnerabilities emphasizes the importance of fully considering the cross-function characteristics when detecting or addressing vulnerabilities.

\section{Limitations of Rule-based Detectors.}\label{cwe_review}
The following presents more analysis of our empirical review of the top CWEs.
\noindent\textbf{CWE-400: Uncontrolled Resource Consumption.}
Uncontrolled Resource Consumption refers to a type of vulnerability where a system fails to properly limit resource usage, leading to exhaustion of system resources such as CPU, memory, disk space, network bandwidth, or file descriptors. 
This can result in performance degradation, denial of service (DoS), or even system crashes.
The examined Uncontrolled Resource Consumption vulnerabilities can be attributed to four causes:
1. Improper limitations on resource consumption \textbf{(23/30)}. A typical example of this includes parsing a user-supplied YAML file without setting the maximum number of nodes, which can
lead to excessive consumption of space or time.
2. Regular expressions with an inefficient worst-case computational complexity \textbf{(4/30)}.
3. Algorithm defects \textbf{(2/30)}. For instance, certain user input can trigger infinite loops in a program.
4. Unclosed resources \textbf{(1/30)}.

Bandit targets only one specific case of improper limitations on resource consumption, which checks whether the \verb|timeout| parameter has been set in the \verb|request| library's API calls, failing to address this most prevalent type of Uncontrolled Resource Consumption systematically. 
Improper limitations on resource consumption attribute to the co-existence of two factors: 
1) User-consumed resources. There exists data flows from user inputs to resource consumption APIs, such as file storage APIs and XML parsing APIs;
2) Absence of limitations or user supplying limitations, such as the size of user-uploaded data for file storage APIs, or the maximum number of nodes in XML parsing APIs. 
These limitations are typically implemented either as parameters of the resource consumption APIs or as independent checks before user inputs reach these APIs.
As such, effective detection of improper limitations on resource consumption requires an extended taint analysis that not only identifies the taint flows from user inputs to resource consumption APIs, but also backtraces the limitations from these APIs.

On the other hand, CodeQL includes a rule targeting inefficient regular expressions, failing to address most Uncontrolled Resource Consumption vulnerabilities, and PySA does not have any rule targeting Uncontrolled Resource Consumption vulnerabilities.

\noindent\textbf{CWE-362: Concurrent Execution using Shared Resource with Improper Synchronization ('Race Condition').}
A race condition can arise when the necessary atomicity of operations is violated in concurrent execution, resulting in unexpected program behavior. Traditional atomicity violations typically involve synchronous operations, such as threads, accessing shared memory without adequate safeguards. In web applications, atomicity violations can also occur when synchronous operations access external resources such as file systems.
In the PyPI ecosystem, both traditional \textbf{(12/30)} and web application-related \textbf{(18/30)} atomicity violations are commonly observed. 

None of the three tools supports detection of race conditions in Python. Detection of traditional atomicity violations involving access to shared memory requires definitions of atomic regions~\cite{zheng2012static}. This detection can potentially be implemented using CodeQL, which provides API modeling based on functionality
and a sound data flow analysis engine. 
Web application-related atomicity violations extend further, requiring an assessment of whether multiple operations access the same external resource, such as a specific data record in a database. 
Furthermore, as discussed, data flows in web applications are complex to model.
As such, detecting atomicity violations in web applications requires sophisticated methods to be developed.



\noindent\textbf{CWE-89: Improper Neutralization of Special Elements used in an SQL Command ('SQL Injection').}
SQL vulnerabilities occur when developers fail to filter, escape, restrict, or properly handle user input strings in systems that interact with databases. 
This allows attackers to input carefully crafted strings to illegally access data from the database.
The majority \textbf{(28/29)} of SQL injection vulnerabilities are caused by improper input validation, except for CVE-2014-0474~\cite{cve20140474}, which mainly relates to developers' unawareness of MySQL's typecasting behavior.

All three detectors target SQL injection caused by improper input validation. Bandit's rules checks for hard-coded SQL queries and use of potentially dangerous APIs such as Django's RawSQL. However, as Bandit does not exhibits any data flow analysis, these rules exhibits a high false positive rate. CodeQL and PySA adopted taint analysis and are able to more accurately identify SQL injection. However, in the vulnerability reports examined, most of these SQL injections locate in non-standalone packages, and taint analysis in these packages are largely ineffective without package-specific taint specifications.

\section{Detailed Clustering}\label{sec:appendix}


\onecolumn
\begin{longtable}{@{}p{14em}p{28em}cc@{}}
\caption{Clustering of CWE vulnerability types}
\label{tab:clustering}
\footnotesize\\
\toprule
\textbf{Cluster   Name}                                                  & \textbf{CWE Name}                                                                                                    & \textbf{Commits} & \textbf{Functions} \\ \midrule
\endfirsthead
\multicolumn{4}{c}%
{{\bfseries Table \thetable\ continued from previous page}} \\
\toprule
\textbf{Cluster   Name}                                                  & \textbf{CWE Name}                                                                                                    & \textbf{Commits} & \textbf{Functions} \\ \hline
\endhead
%
\hline
\endfoot
\endlastfoot
\multirow[t]{17}{*}{Injection}                                     & CWE-79 Improper Neutralization of Input During Web Page Generation   ('Cross-site Scripting')               & 89      & 185       \\
                                                                & CWE-89 Improper Neutralization of Special Elements used in an SQL Command ('SQL Injection')                 & 29      & 50        \\
                                                                & CWE-78 Improper Neutralization of Special   Elements used in an OS Command ('OS Command Injection')         & 20      & 29        \\
                                                                & CWE-74 Improper Neutralization of Special   Elements in Output Used by a Downstream Component ('Injection') & 12      & 29        \\
                                                                & CWE-94 Improper Control of Generation of   Code ('Code Injection')                                          & 12      & 36        \\
                                                                & CWE-77 Improper Neutralization of Special   Elements used in a Command ('Command Injection')                & 10      & 11        \\
                                                                & CWE-611 Improper Restriction of XML   External Entity Reference                                             & 7       & 17        \\
                                                                & CWE-88 Improper Neutralization of   Argument Delimiters in a Command ('Argument Injection')                 & 6       & 11        \\
                                                                & CWE-1336 Improper Neutralization of   Special Elements Used in a Template Engine                            & 4       & 11        \\
                                                                & CWE-93 Improper Neutralization of CRLF   Sequences ('CRLF Injection')                                       & 3       & 11        \\
                                                                & CWE-80 Improper Neutralization of   Script-Related HTML Tags in a Web Page (Basic XSS)                      & 3       & 5         \\
                                                                & CWE-116 Improper Encoding or Escaping of   Output                                                           & 2       & 4         \\
                                                                & CWE-75 Failure to Sanitize Special   Elements into a Different Plane (Special Element Injection)            & 1       & 3         \\
                                                                & CWE-707 Improper Neutralization                                                                             & 1       & 1         \\
                                                                & CWE-1236 Improper Neutralization of   Formula Elements in a CSV File                                        & 1       & 2         \\
                                                                & CWE-96 Improper Neutralization of   Directives in Statically Saved Code ('Static Code Injection')           & 1       & 4         \\
                                                                & CWE-91 XML Injection (aka Blind XPath   Injection)                                                          & 1       & 2         \\ \hline
\multirow[t]{2}{*}{Improper Input   Validation}                    & CWE-20 Improper Input Validation                                                                            & 69      & 99        \\
                                                                & CWE-1284 Improper Validation of Specified   Quantity in Input                                               & 6       & 10        \\ \hline
\multirow[t]{8}{*}{File Operation   Error}                         & CWE-22 Improper Limitation of a Pathname to a Restricted Directory ('Path   Traversal')                     & 51      & 101       \\
                                                                & CWE-59 Improper Link Resolution Before   File Access ('Link Following')                                     & 10      & 20        \\
                                                                & CWE-377 Insecure Temporary File                                                                             & 5       & 6         \\
                                                                & CWE-434 Unrestricted Upload of File with   Dangerous Type                                                   & 4       & 19        \\
                                                                & CWE-29 Path Traversal:   '\textbackslash{}..\textbackslash{}filename'                                       & 4       & 10        \\
                                                                & CWE-23 Relative Path Traversal                                                                              & 4       & 6         \\
                                                                & CWE-36 Absolute Path Traversal                                                                              & 1       & 1         \\
                                                                & CWE-641 Improper Restriction of Names for   Files and Other Resources                                       & 1       & 2         \\ \hline
NULL Pointer Dereference                                        & CWE-476 NULL Pointer Dereference                                                                            & 43      & 53        \\ \hline
\multirow[t]{6}{*}{Out-of-Bound   Read/Write}                      & CWE-125 Out-of-bounds Read                                                                                  & 43      & 53        \\
                                                                & CWE-787 Out-of-bounds Write                                                                                 & 21      & 48        \\
                                                                & CWE-119 Improper Restriction of   Operations within the Bounds of a Memory Buffer                           & 17      & 19        \\
                                                                & CWE-120 Buffer Copy without Checking Size   of Input ('Classic Buffer Overflow')                            & 16      & 28        \\
                                                                & CWE-131 Incorrect Calculation of Buffer   Size                                                              & 11      & 17        \\
                                                                & CWE-122 Heap-based Buffer Overflow                                                                          & 6       & 9         \\ \hline
Resource Management Error                & CWE-400 Uncontrolled Resource Consumption                                                                   & 41      & 78        \\
                                                                & CWE-770 Allocation of Resources Without   Limits or Throttling                                              & 12      & 31        \\
                                                                & CWE-404 Improper Resource Shutdown or   Release                                                             & 1       & 3         \\ \hline
Assertion Failures                                              & CWE-617 Reachable Assertion                                                                                 & 39      & 46        \\ \hline
\multirow[t]{11}{*}{Information   Exposure}                        & CWE-200 Exposure of Sensitive Information to an Unauthorized Actor                                          & 38      & 64        \\
                                                                & CWE-209 Generation of Error Message   Containing Sensitive Information                                      & 4       & 7         \\
                                                                & CWE-532 Insertion of Sensitive   Information into Log File                                                  & 4       & 5         \\
                                                                & CWE-212 Improper Removal of Sensitive   Information Before Storage or Transfer                              & 4       & 11        \\
                                                                & CWE-312 Cleartext Storage of Sensitive   Information                                                        & 2       & 6         \\
                                                                & CWE-668 Exposure of Resource to Wrong   Sphere                                                              & 2       & 3         \\
                                                                & CWE-614 Sensitive Cookie in HTTPS Session   Without 'Secure' Attribute                                      & 2       & 2         \\
                                                                & CWE-598 Use of GET Request Method With   Sensitive Query Strings                                            & 1       & 1         \\
                                                                & CWE-524 Use of Cache Containing Sensitive   Information                                                     & 1       & 1         \\
                                                                & CWE-213 Exposure of Sensitive Information   Due to Incompatible Policies                                    & 1       & 2         \\
                                                                & CWE-311 Missing Encryption of Sensitive   Data                                                              & 1       & 1         \\ \hline
\multirow[t]{5}{*}{Incorrect   Synchronization}                    & CWE-362 Concurrent Execution using Shared Resource with Improper   Synchronization ('Race Condition')       & 34      & 62        \\
                                                                & CWE-367 Time-of-check Time-of-use   (TOCTOU) Race Condition                                                 & 1       & 1         \\
                                                                & CWE-821 Incorrect Synchronization                                                                           & 1       & 1         \\
                                                                & CWE-662 Improper Synchronization                                                                            & 1       & 1         \\
                                                                & CWE-833 Deadlock                                                                                            & 1       & 1         \\ \hline
Open Redirect                                                   & CWE-601 URL Redirection to Untrusted Site ('Open Redirect')                                                 & 23      & 47        \\ \hline
Improper Deserialization                                        & CWE-502 Deserialization of Untrusted Data                                                                   & 23      & 39        \\ \hline
\multirow[t]{4}{*}{Origin Validation   Error}                      & CWE-352 Cross-Site Request Forgery (CSRF)                                                                   & 23      & 54        \\
                                                                & CWE-918 Server-Side Request Forgery   (SSRF)                                                                & 21      & 36        \\
                                                                & CWE-444 Inconsistent Interpretation of   HTTP Requests ('HTTP Request/Response Smuggling')                  & 9       & 23        \\
                                                                & CWE-346 Origin Validation Error                                                                             & 1       & 1         \\ \hline
\multirow[t]{34}{*}{Improper  Access Control}                      & CWE-287 Improper Authentication                                                                             & 20      & 38        \\
                                                                & CWE-284 Improper Access Control                                                                             & 14      & 28        \\
                                                                & CWE-863 Incorrect Authorization                                                                             & 10      & 18        \\
                                                                & CWE-347 Improper Verification of   Cryptographic Signature                                                  & 10      & 30        \\
                                                                & CWE-295 Improper Certificate   Validation                                                                   & 9       & 34        \\
                                                                & CWE-384 Session Fixation                                                                                    & 7       & 32        \\
                                                                & CWE-522 Insufficiently Protected   Credentials                                                              & 6       & 10        \\
                                                                & CWE-285 Improper Authorization                                                                              & 5       & 12        \\
                                                                & CWE-276 Incorrect Default   Permissions                                                                     & 5       & 7         \\
                                                                & CWE-269 Improper Privilege   Management                                                                     & 4       & 14        \\
                                                                & CWE-345 Insufficient Verification of Data   Authenticity                                                    & 4       & 4         \\
                                                                & CWE-640 Weak Password Recovery Mechanism   for Forgotten Password                                           & 4       & 4         \\
                                                                & CWE-294 Authentication Bypass by   Capture-replay                                                           & 3       & 4         \\
                                                                & CWE-250 Execution with Unnecessary   Privileges                                                             & 3       & 4         \\
                                                                & CWE-307 Improper Restriction of Excessive   Authentication Attempts                                         & 3       & 8         \\
                                                                & CWE-521 Weak Password Requirements                                                                          & 3       & 3         \\
                                                                & CWE-290 Authentication Bypass by   Spoofing                                                                 & 2       & 2         \\
                                                                & CWE-306 Missing Authentication for   Critical Function                                                      & 2       & 7         \\
                                                                & CWE-862 Missing Authorization                                                                               & 2       & 5         \\
                                                                & CWE-1220 Insufficient Granularity of   Access Control                                                       & 2       & 6         \\
                                                                & CWE-620 Unverified Password Change                                                                          & 2       & 12        \\
                                                                & CWE-305 Authentication Bypass by Primary   Weakness                                                         & 1       & 4         \\
                                                                & CWE-289 Authentication Bypass by   Alternate Name                                                           & 1       & 2         \\
                                                                & CWE-288 Authentication Bypass Using an   Alternate Path or Channel                                          & 1       & 4         \\
                                                                & CWE-304 Missing Critical Step in   Authentication                                                           & 1       & 2         \\
                                                                & CWE-639 Authorization Bypass Through   User-Controlled Key                                                  & 1       & 1         \\
                                                                & CWE-273 Improper Check for Dropped   Privileges                                                             & 1       & 1         \\
                                                                & CWE-613 Insufficient Session   Expiration                                                                   & 1       & 1         \\
                                                                & CWE-749 Exposed Dangerous Method or   Function                                                              & 1       & 1         \\
                                                                & CWE-940 Improper Verification of Source   of a Communication Channel                                        & 1       & 2         \\
                                                                & CWE-281 Improper Preservation of   Permissions                                                              & 1       & 1         \\
                                                                & CWE-732 Incorrect Permission Assignment   for Critical Resource                                             & 1       & 1         \\
                                                                & CWE-942 Permissive Cross-domain Policy   with Untrusted Domains                                             & 1       & 2         \\
                                                                & CWE-322 Key Exchange without Entity   Authentication                                                        & 1       & 1         \\ \hline
\multirow[t]{6}{*}{Computation Error}                              & CWE-369 Divide By Zero                                                                                      & 36      & 38        \\
                                                                & CWE-190 Integer Overflow or   Wraparound                                                                    & 19      & 24        \\
                                                                & CWE-681 Incorrect Conversion between   Numeric Types                                                        & 6       & 9         \\
                                                                & CWE-191 Integer Underflow (Wrap or   Wraparound)                                                            & 2       & 2         \\
                                                                & CWE-682 Incorrect Calculation                                                                               & 2       & 5         \\
                                                                & CWE-193 Off-by-one Error                                                                                    & 1       & 1         \\ \hline
\multirow[t]{2}{*}{Regular Expression}                             & CWE-1333 Inefficient Regular Expression Complexity                                                          & 18      & 25        \\
                                                                & CWE-185 Incorrect Regular Expression                                                                        & 4       & 6         \\ \hline
\multirow[t]{4}{*}{Uncontrolled   Recursion}                       & CWE-674 Uncontrolled Recursion                                                                              & 4       & 5         \\
                                                                & CWE-835 Loop with Unreachable Exit   Condition ('Infinite Loop')                                            & 10      & 17        \\
                                                                & CWE-776 Improper Restriction of Recursive   Entity References in DTDs ('XML Entity Expansion')              & 1       & 1         \\
                                                                & CWE-834 Excessive Iteration                                                                                 & 1       & 1         \\ \hline
\multirow[t]{3}{*}{Uninitialized}                                  & CWE-824 Access of Uninitialized Pointer                                                                     & 10      & 11        \\
                                                                & CWE-665 Improper Initialization                                                                             & 8       & 8         \\
                                                                & CWE-908 Use of Uninitialized   Resource                                                                     & 6       & 10        \\ \hline
\multirow[t]{5}{10em}{Improper Exception handling}                  & CWE-754 Improper Check for Unusual or Exceptional Conditions                                                & 9       & 12        \\
                                                                & CWE-12 ASP.NET Misconfiguration: Missing   Custom Error Page                                                & 3       & 4         \\
                                                                & CWE-755 Improper Handling of Exceptional   Conditions                                                       & 2       & 2         \\
                                                                & CWE-460 Improper Cleanup on Thrown   Exception                                                              & 1       & 1         \\
                                                                & CWE-248 Uncaught Exception                                                                                  & 1       & 1         \\ \hline
\multirow[t]{3}{*}{Incomplete   Cleanup}                           & CWE-459 Incomplete Cleanup                                                                                  & 2       & 4         \\
                                                                & CWE-416 Use After Free                                                                                      & 2       & 4         \\
                                                                & CWE-415 Double Free                                                                                         & 2       & 3         \\ \hline
\multirow[t]{3}{*}{Side Channel}                                   & CWE-203 Observable Discrepancy                                                                              & 2       & 4         \\
                                                                & CWE-385 Covert Timing Channel                                                                               & 2       & 3         \\
                                                                & CWE-208 Observable Timing   Discrepancy                                                                     & 1       & 1         \\ \hline
Format String                                                   & CWE-134 Use of Externally-Controlled Format String                                                          & 2       & 4         \\ \hline
\multirow[t]{7}{10em}{Inefficient   Algorithmic Complexity}           & CWE-330 Use of Insufficiently Random Values                                                                 & 1       & 3         \\
                                                                & CWE-331 Insufficient Entropy                                                                                & 2       & 5         \\
                                                                & CWE-338 Use of Cryptographically Weak   Pseudo-Random Number Generator (PRNG)                               & 1       & 1         \\
                                                                & CWE-328 Use of Weak Hash                                                                                    & 1       & 1         \\
                                                                & CWE-407 Inefficient Algorithmic   Complexity                                                                & 2       & 2         \\
                                                                & CWE-326 Inadequate Encryption   Strength                                                                    & 3       & 21        \\
                                                                & CWE-327 Use of a Broken or Risky   Cryptographic Algorithm                                                  & 2       & 6         \\ \hline
Incorrect Provision of Specified Functionality                & CWE-684 Incorrect Provision of Specified Functionality                                                      & 9       & 27        \\ \hline
Always-Incorrect Control Flow Implementation                  & CWE-670 Always-Incorrect Control Flow Implementation                                                        & 5       & 7         \\ \hline
Improper Validation of Integrity Check Value                  & CWE-354 Improper Validation of Integrity Check Value                                                        & 5       & 8         \\ \hline
Incorrect Comparison                                            & CWE-697 Incorrect Comparison                                                                                & 4       & 10        \\ \hline
Incorrect Type Conversion or Cast                             & CWE-704 Incorrect Type Conversion or Cast                                                                   & 2       & 2         \\ \hline
Improper Handling of Alternate Encoding                       & CWE-173 Improper Handling of Alternate Encoding                                                             & 2       & 6         \\ \hline
Improper Handling of Structural Elements                      & CWE-237 Improper Handling of Structural Elements                                                            & 2       & 2         \\ \hline
Business Logic Errors                                           & CWE-840 Business Logic Errors                                                                               & 2       & 6         \\ \hline
Acceptance of Extraneous Untrusted Data With Trusted Data     & CWE-349 Acceptance of Extraneous Untrusted Data With Trusted Data                                           & 2       & 2         \\ \hline
Access of Resource Using Incompatible Type ('Type Confusion') & CWE-843 Access of Resource Using Incompatible Type ('Type   Confusion')                                     & 2       & 6         \\ \hline
Unprotected Alternate   Channel                                 & CWE-420 Unprotected Alternate Channel                                                                       & 2       & 2         \\ \hline
Undefined Behavior for Input to API                           & CWE-475 Undefined Behavior for Input to API                                                                 & 2       & 2         \\ \hline
Prototype Pollution                                             & CWE-1321 Improperly Controlled Modification of Object Prototype   Attributes ('Prototype Pollution')        & 1       & 1         \\ \hline
Improper Output Neutralization   for Logs                       & CWE-117 Improper Output Neutralization for Logs                                                             & 1       & 2         \\ \hline
Client-Side Enforcement of Server-Side Security               & CWE-602 Client-Side Enforcement of Server-Side Security                                                     & 1       & 2         \\ \hline
Improper Restriction of Rendered UI Layers or Frames          & CWE-1021 Improper Restriction of Rendered UI Layers or Frames                                               & 1       & 1         \\ \hline
Unchecked Return Value                                          & CWE-252 Unchecked Return Value                                                                              & 1       & 1         \\ \hline
Initialization of a Resource with an Insecure Default         & CWE-1188 Initialization of a Resource with an Insecure Default                                              & 1       & 1         \\ \hline
Mutable Attestation or Measurement Reporting Data             & CWE-1283 Mutable Attestation or Measurement Reporting Data                                                  & 1       & 1         \\ \hline
Function Call With Incorrect Order of Arguments               & CWE-683 Function Call With Incorrect Order of Arguments                                                     & 1       & 1         \\ \hline
Interpretation Conflict                                         & CWE-436 Interpretation Conflict                                                                             & 1       & 1         \\ \hline
Improper Control of Dynamically-Managed Code Resources        & CWE-913 Improper Control of Dynamically-Managed Code Resources                                              & 1       & 1         \\ \hline
Unimplemented or \\Unsupported Feature in UI                   & CWE-447 Unimplemented or Unsupported Feature in UI                                                          & 1       & 1         \\ \bottomrule
\end{longtable}
\end{document}